\documentclass{aa}

\usepackage{graphicx}
\usepackage{txfonts}
\usepackage{natbib}
\bibpunct{(}{)}{;}{a}{}{,} 
\def\arcm{\hbox{$^\prime$}}
\def\arcs{\arcm\hskip -0.1em\arcm}

\def\ergs{erg s$^{-1}$\xspace}

\newcommand\sloanr{$r^\prime$}

\begin{document}

   \title{A census of H$\alpha$ emitters in the intergalactic medium of the NGC~2865 system\thanks{Based on observations obtained at the Gemini Observatory, which is operated by the Association of Universities for Research in Astronomy, Inc.,  under a cooperative agreement with the NSF on behalf of the Gemini partnership: the National Science Foundation (United States), the Science and Technology Facilities Council (United Kingdom), the National Research Council (Canada), CONICYT (Chile), the Australian Research Council (Australia), Minist\'erio da Ci\^encia e Tecnologia (Brazil) and Ministerio de Ciencia, Tecnolog\'ia e Innovaci\'on Productiva (Argentina) - Observing runs: GS-2008A-Q-35.}}

   \subtitle{}

   \author{F. Urrutia-Viscarra 
          \inst{1,}\inst{2}
          \and
          M. Arnaboldi\inst{1}
          \and
          C. Mendes de Oliveira\inst{2}          
          \and
          O. Gerhard\inst{3}
          \and
          S. Torres-Flores\inst{4}
          \and
          E. R. Carrasco\inst{5}
          \and
          D. de Mello\inst{6,7}
          }

   \institute{European Southern Observatory, Karl-Schwarzschild-Strasse 2, 85748 Garching, Germany         
         \and
             Instituto de Astronomia, Geofísica e Ciências Atmosféricas, Universidade de São Paulo, Rua do Matão 1226,
Cidade Universitária, 05508–900, São Paulo, SP, Brazil \email{furrutia@astro.iag.usp.br}
		 \and
		 Max-Planck-Institut f\"ur Extraterrestrische Physik, Giessenbachstrasse, D-85741 Garching, Germany
		 \and
		 Departamento de Física, Universidad de La Serena, Av. Cisternas 1200 Norte, La Serena, Chile
		 \and
		 Gemini Observatory/AURA, Southern Operations Center, Casilla 603, La Serena, Chile 
		 \and
		 Physics Department, The Catholic University of America, Washington, DC 20064 USA
		 \and
		 Observational Cosmology Laboratory, Code 665, Goddard Space Flight Center, Greenbelt, MD 20771 USA
             }

   \date{Received , 2013; accepted }

 
  \abstract
   { Tidal debris which are rich in H{\sc I} gas, formed in interacting and merging systems, are suitable laboratories to study star formation outside galaxies. Recently, several such systems were observed, which contained many young star forming regions outside the galaxies.} 
   { In previous works, we have studied young star forming regions outside galaxies in different systems with optical and/or gaseous tidal debris, all of them with available archive GALEX/UV images, in order to understand how often they occur and in which type of environments. In this paper we searched for star forming regions around the galaxy NGC~2865, a shell galaxy which is circled by a ring of H{\sc I}, with a total mass of 1.2 $\times$ 10$^9$ M$_\odot$.}
   { Using the Multi-Slit Imaging Spectroscopy Technique with the Gemini telescope, we detected all H$\alpha$ emitting sources in the surroundings of the galaxy NGC~2865, down to a flux limit of 10$^{-18}$ erg cm$^{-2}$ s$^{-1}$ \AA$^{-1}$. Together with Near and Far-Ultraviolet flux information we characterize the star formation rates, masses, ages, and metallicities for these H{\sc II} regions. In total, we found 26 emission-line sources in a 60 $\times$ 60 Kpc field centered over the southeastern tail of the H{\sc I} gas present around the galaxy NGC~2865.}
   { Out of the 26 H$\alpha$ emitters, 19 are in the satellite galaxy FGCE 0745 and seven are intergalactic H{\sc II} regions scattered over the south tail of the H{\sc I} gas around NGC~2865. We found that the intergalactic H{\sc II} regions are young ($<$200 Myr)  with stellar masses in the range 4$\times$10$^3$M$_\odot$  to 17$\times$10$^6$ M$_\odot$. These are found in a region of low H{\sc I} gas density, where the probability of forming stars is expected to be low. For one of the intergalactic H{\sc II} regions we estimated a solar oxygen abundance of 12 + log(O/H) $\sim$ 8.7. We also were able to estimate the metallicity for the satellite galaxy FGCE~0745 to be 12 + log(O/H) $\sim$ 8.0.}
   {Given these physical parameters, the intergalactic H{\sc II} regions are consistent with young star forming regions (or clusters), born in situ outside the NGC~2865 galaxy from a pre-enriched gas removed from the host galaxies in a merger event. The relevance of these observations is discussed.}
    

   \maketitle

\section{Introduction}

Stars are formed in dense regions inside giant molecular clouds. However, what triggers and quenches the cloud collapse is still an open question that has strong implications for the understanding of how galaxies form and evolve. One approach to this problem is the study of star formation in extreme environments. One of these environments is the tidal debris of H{\sc I} gas caused by galaxy-galaxy interactions and galaxy mergers. They are excellent candidates to contain a large number of H{\sc II} regions outside galaxies, as illustrated by the numerical simulations of \citet{Bournaud08}. With the detections of these intergalactic systems we can analyze the probability to form stellar systems in environments where the gas density is very low, H{\sc I} $\sim$ 10$^{19}$ cm$^{-2}$. In recent years various authors reported intergalactic objects in interacting systems with gas tails, such as: tidal dwarf galaxies \citep[e.g.][]{Duc98,Duc00,mendes01,Hibbard01,demello12,Lee12} and intergalactic H{\sc II} regions \citep[e.g.][]{Gerhard02,ryan04,mendes04,Arrigoni12,torres12,demello12,Yagi13}.

 When one or more gas rich galaxies are involved in an interaction, it is common to find extended tails of H{\sc I} gas outside these galaxies. Galaxy interactions are rare phenomena in the local Universe, and even rarer if the interaction involves two massive galaxies, i.e. a major merger. While these represent some of the most spectacular collisions we observe, they are far from being common. Because of the shape of the galaxy luminosity function, which rises at fainter luminosities, encounters more commonly involve a bright galaxy interacting with a small satellite. The optical Hubble Space Telescope Deep Sky Survey shows that the number of galaxies with signatures of interaction increase by approximately 10$\%$ between redshift 0.7 and 1 \citep[e.g.][]{Fevre00,Lopez13} compared to the Local Universe. One of the environmental effects triggered by galaxy interactions is extended star formation. This is commonly observed in pairs or in close interacting galaxies, mainly in the systems which contain disturbed H{\sc I} features. Recently, many efforts have been devoted to several systems in the local universe to understand the physical conditions that lead to the formation of new stars and whether these are responsible for polluting the intergalactic medium with metal-enriched gas \citep[e.g.][]{mendes04,ryan04,Bournaud04,demello12,torres12,Arrigoni12,Yagi13}. These are important points, given that many studies \citep[i.e.][]{Songaila01,Pettini03,Becker06,Ryan06,Ryan09,Simcoe11,Diaz11,Dodorico13} showed that the intergalactic medium at redshift at least 6 is not a pristine remnant of the Big Bang. But rather it contains significant quantities of metals.

In this paper we study the surroundings of the elliptical galaxy NGC~2865, which has an unusual quantit of H{\sc I} gas around it \citep[M$_{H{\sc I}}$ $\sim$  1.2$\times$10$^9$ M$_\odot$,][]{schiminovich95}. NGC~2865 is classified as an E3 galaxy in the RC3 catalog \citep{Vaucouleurs91} and it satisfies the \citet{faber76} relation \citep{lake86}. However, deep images of NGC~2865 demonstrate that it is a genuinely peculiar galaxy. The galaxy clearly shows a significantly disturbed morphology. \citet{malin83} cataloged NGC~2865 as a shell galaxy: an external shell at about 2\arcmin\ east of the nucleus is barely visible both in the NUV and in the HST ACS image. Also, a faint loop is visible to the northwest of the galaxy while a tail extension can be seen to the southeast \citep{Rampazzo07}. These two features are typically associated with galaxy interactions. Stellar spectroscopy and UBV photometry set lower and upper limits to the age of the possible encounter to 1 and 4 Gyr respectively, consistent with major merger models \citep{schiminovich95}. This encounter would have lead to the formation of NGC~2865. In fact, stellar spectroscopy of the nuclear regions of NGC~2865 \citep{bica87} shows a bump in the spectrum at 4600 \AA\ and a corresponding strengthening of the Balmer lines, features that may correspond to an intermediate-age burst of star formation. In addition, NGC~2865 presents an unusual ring of H{\sc I} gas around it, that may be a remnant of a merger event. NGC~2865 has no nearby galaxies of similar luminosity, only two gas-rich  galaxies are seen nearby, FGCE 0745 and [M98k] 092035.0-225654, respectively 6\arcmin\ and 9\arcmin\ from NGC~2865, i.e. 1.0 and 1.5 kpc . Basic physical parameters for NGC~2865 and the nearest galaxy FGCE 0745 are listed in Table \ref{table:literature}, M98k has not been observed in this work. Other authors studying similar systems (considering the morphological types of the galaxies) are \citet{Boselli05} for the interacting pairs M~86/NGC~4438 and \citet{Arrigoni12} for the system VCC1249/M49, where several intergalactic H{\sc II} regions were discovered. 

In this article, we address the question whether star formation is possible in low-density H{\sc I} gas and if  the H{\sc I} gas around NGC~2865 is primordial. If the gas is not primordial, it must be removed from a gas-rich galaxy by tidal effect during an interaction and now is in the intergalactic medium.

The layout of this paper is as follows. In Section 2 the novel observational technique and the data reduction are described. Data analysis is shown  in Section 3. In Section 4  the principal properties and the results for each H{\sc II} region are presented.  This is followed by a discussion in Section 5 and the conclusion in Section 6. Throughout the paper, we assume $\Omega_M$=0.3, $\Omega_\Lambda$=0.7 and H$_0$=100 $h$ km s$^{-1}$ Mpc$^{-1}$, with $h$ = 0.75.

\begin{table}
\caption{Principal parameters for NGC~2865 and the satellite galaxy FGCE~0745, from the literature.}             
\label{table:literature}      
\centering      
\small
\tabcolsep=0.1cm
\begin{tabular}{c c c c c c c}    
\hline\hline       
 & V$_{sys}$\tablefootmark{a} &  L$_B$           & r$_{e}$    & $\sigma_C$     & M$_{HI}$      & V$_{HI}$  \\
 &    \tiny{Km s$^{-1}$}      &  L$_\odot$       &   kpc      &  km s$^{-1}$    & M$_\odot$     & Km s$^{-1}$                          \\ 
\hline                    
NGC~2865  & 2627$\pm$3 & 2.4$\times$10$^{10}$\tablefootmark{c}  & 1.33\tablefootmark{c} & 200\tablefootmark{b}  & 1.2$\times$10$^8$\tablefootmark{c}  & 2694$\pm$15\tablefootmark{c}  \\
FGCE 0745 & 2480$\pm$14  & ---  & --- & ---  & 4.9$\times$10$^8$  & 2725\tablefootmark{c}  \\
\hline                  
\end{tabular}
\tablefoot{
\tablefoottext{a}{\citep{Smith00}, NED}
\tablefoottext{b}{The central velocity dispersion $\sigma_C$ was taken from \citet{Hau99}}
\tablefoottext{c}{Taken from Schiminovich et al. 1995}
}
\end{table}

\section{Observations}

\subsection{Spectroscopic Observations}

The data were collected with the Gemini Multi-Object Spectrograph (Hook et al. 2004, hereafter GMOS) mounted on the Gemini South telescope in Chile in queue mode (Program ID. GS-2008A-Q-35). 

The southeastern  HI tail of NGC~2865 ($\alpha(2000)=$ 9$^{h}$ 23$^{m}$ 37\fs13, $\delta(2000)=$ $-23$\degr 11\arcmin\ 54\farcs34) was imaged with the \sloanr\ filter on 2008 Jan 28 (UT) in clear conditions and with an average seeing of 0.95\arcsec. The images were processed in the standard manner with the Gemini IRAF package (version 1.8). The final combined images were used to build the multi-slit mask.

\begin{figure}
   \centering
      \includegraphics[width=\hsize]{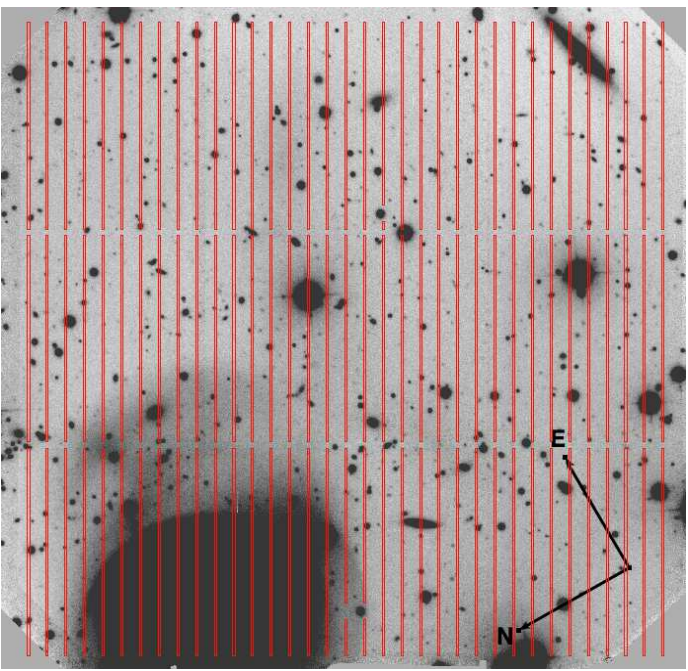}
      \caption{The mask overlaid onto the \sloanr-band pre-imaging of 5.5\arcmin\ $\times$ 5.5\arcmin. The slits are arranged in a total of 108 long slits with two short interruptions for mechanical stability of the mask. The width of each slit is 1 arcsec. This position mask is for POS0.}
         \label{mask}
   \end{figure}

A technique called Multi-Slit Imaging Spectroscopy  (MSIS) has been used to search for HII regions in the gaseous tail of NGC~2865. The use of the MSIS technique requires the construction of a special mask with multiple long slits spaced by few arcsecs combined with a narrow-band filter. We built a mask with multiple parallel long slits of 1\arcsec in width and spaced every 8\arcsec. Figure~\ref{mask} shows the combined \sloanr\ image with the mask overlaid onto it.  

The spectra were observed on 2008 Apr 13 (UT), during dark time, through thin cirrus and with a seeing that varied between 0\farcs6 and 0\farcs8. The mask was observed in 9 different positions, each one shifted spatially by 1\arcsec with respect to the previous one, always in the same direction. The dither pattern used allowed to cover $\sim 95$\% of the field inside the 5\arcmin $\times$ 5\arcmin\ GMOS field of view. At each position (nine in total),  3$\times$315 sec exposures were obtained with the 400 lines mm$^{-1}$ ruling  density grating (R400), centered at 6620\AA\ and using the H$\alpha$ continuum filter \footnote[1]{http://www.gemini.edu/node/10637}. Internal flats were obtained every 3 science exposures and CuAr (copper-argon) arc lamp calibration frames were acquired at the beginning and the end of the observing sequence. In addition, the CuAr arc lamp  was observed without the H$\alpha$ continuum filter at the end of the observing sequence. The selected instrument setup (grating $+$ narrow band filter $+$ central wavelength) has been chosen to optimize the detection of any H$\alpha$ emissions for all emitting objects at the rest frame of the NGC~2865 galaxy down to a flux of 10$^{-18}$ \ergs\ cm$^{-2}$ \AA$^{-1}$. At the redshift of the galaxy, the H$\alpha$ lines are expected to be around 6563\AA. The log of the observations is given in Table \ref{logbook}.

All spectra were bias subtracted, trimmed, flat fielded, and wavelength calibrated using the Gemini IRAF package version 1.8. To calibrate the spectra in wavelength, we first used the CuAr lamp observed without the H$\alpha$ continuum filter to identify the lines belonging to the wavelength region covered by our observations (6590\AA\ - 6660\AA) and to asses the errors associated to the wavelength calibration. We used the Argon lines at 6604\AA\ and 6643\AA, which are the two strongest lines visible, to calibrate the spectra. Using the CuAr lamp observed without the H$\alpha$ continuum filter, the residual values in the wavelength solution for 70-77 points using a 4th-order Chebyshev polynomial typically yielded rms values of $\sim$ 0.8\AA. Finally, the spectra were flux calibrated using the spectrophotometric standard star LTT 9239. The final spectra have a wavelength coverage between $\sim$ 6590\AA\ and $\sim$ 6660\AA\  ($\sim$ 70\AA), an instrumental resolution of $\sim 9$\AA\ at 6620\AA, and a dispersion of $\sim 1.5$\AA/pixel. 

\begin{table}
\caption{Observation logbook.}             
\label{logbook}      
\centering      
\small
\tabcolsep=0.1cm
\begin{tabular}{c c c c c c c}   
\hline\hline       
Date & Mask & Total\tablefootmark{a} & Average & BinX     & Number of & Xoffset\tablefootmark{b} \\
     &      & exposure               & seeing  & $\times$ & frames    & \\ 
     &      & sec.                   &         & BinY     &           & \\ 
\hline                    
13.04.2008 & POS0  & 315  & 0\farcs7 & 2$\times$2  & 3 &  0\farcs73  \\
13.04.2008 & POS1  & 315  & 0\farcs7 & 2$\times$2  & 3 & -0\farcs27  \\
13.04.2008 & POS2  & 315  & 0\farcs6 & 2$\times$2  & 3 & -1\farcs27  \\
13.04.2008 & POS3  & 315  & 0\farcs7 & 2$\times$2  & 3 & -2\farcs27  \\
13.04.2008 & POS4  & 315  & 0\farcs8 & 2$\times$2  & 3 & -3\farcs27  \\
13.04.2008 & POS5  & 315  & 0\farcs7 & 2$\times$2  & 3 & -4\farcs27      \\
13.04.2008 & POS6  & 315  & 0\farcs7 & 2$\times$2  & 3 & -5\farcs27   \\
13.04.2008 & POS7  & 315  & 0\farcs7 & 2$\times$2  & 3 & -6\farcs27     \\
13.04.2008 & POS8  & 315  & 0\farcs8 & 2$\times$2  & 3 & -7\farcs27  \\
\hline                  
\end{tabular}
\tablefoot{
\tablefoottext{a}{Total exposure time for each frame.}
\tablefoottext{b}{The field of view was centered at $\alpha$(J2000) 09$^{h}$ 23$^{m}$ 37\fs13 and $\delta$(J2000)-23\degr\ 11\arcmin\ 54\farcs34.}
}
\end{table}

\subsubsection{The advantages of the MSIS Technique}

 MSIS is a novel technique which is very efficient at
  searching for faint H{\sc II} regions compared to standard narrow
  band imaging. The MSIS technique has been successfully used to detect
  planetary nebulae
  \citep[e.g.][]{gerhard05,gerhard07,arnaboldi07,ventimiglia11}. This
  is the first time that this technique is applied to carry out a
  survey of H{\sc II} regions in the external regions of a
  galaxy. Unlike previous studies
  \citep[e.g.][]{mendes04,ryan04,demello12,Lee12},the MSIS is a blind
  technique that is capable of finding all the H$\alpha$ emissions in
  a field of view of few arcminutes squared, down to a certain flux limit in
  the H$\alpha$ line of 10$^{-18}$ erg cm$^{-2}$ s$^{-1}$
  \AA$^{-1}$. Thus we are able to do a complete census of H{\sc II}
  regions in the surveyed field. In narrow band imaging, detections
  are sky limited by the sky noise from 100 \AA, which is typically
  the FWHM of H$\alpha$ narrow band filter. In the MSIS, the noise
  from the sky comes from few \AA\ only, depending on the slit width
  and seeing. Hence fluxes of order 10 time fainter can be detected.

\subsubsection{H$\alpha$ candidates.}

At each mask position, the average MSIS frames were inspected for the presence of emission line objects. Here we found three kinds of emission line objects:
\begin{itemize} 

\item  Resolved/unresolved, both in wavelength and space, emission line objects with high equivalent widths, which are the H{\sc II} region candidates. We consider a source to be resolved in wavelength if the full width at half maximum (FWHM) of the lines in its spectrum is larger than the FWHM measured for the typical ARC lamp line. A source is spatially resolved if the FWHM in the y-direction is larger than the seeing.

\item  Continuum sources with line emission. These are spectra from the satellite galaxy FGCE 0745, with an average radial velocity of $\sim$ 2878 km s$^{-1}$.

\item  Continuum sources without line emission. These spectra are
background galaxies or field stars.

\end{itemize}

   \begin{figure}
   \centering
      \includegraphics[width=5cm]{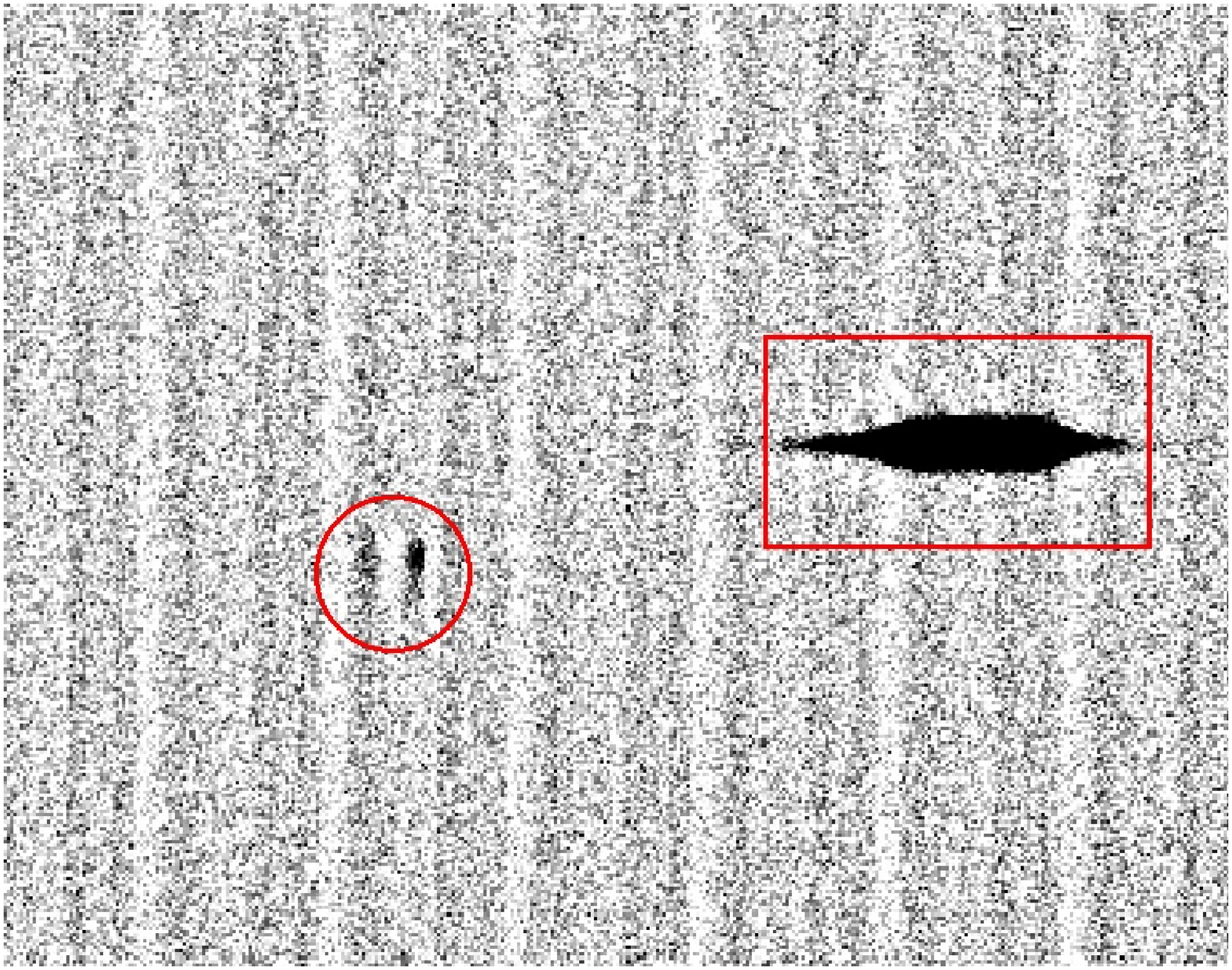}
      \caption{Two-dimensional spectra of emission objects in the Gemini Field, position 4. The spatial direction goes along the y-axis. The wavelength goes along the horizontal axis (6580 - 6670\AA); the spectral resolution is 9\AA{}, or 411 km s$^{-1}$. Each vertical stripe represent a spectrum of 68\AA{} wide. The circle marks the region IG\_04\_P4 with a flux of 3.2$\times$10$^{-16}$ erg s$^{-1}$ cm$^{-2}$. The rectangle shows the spectrum of a background galaxy with continuum.}
         \label{2D}
   \end{figure}
   
A section of the  average MSIS two-dimensional spectrum for the mask position 4 (POS4) is shown in Figure~\ref{2D}. Each vertical stripe in Figure~\ref{2D} represents a spectrum $\sim$ 70\AA{} wide. The spatial direction is along the y-axis, while the wavelength direction for each of the $\sim$ 70\AA{}-wide spectra is along the horizontal x-axis.  In Figure~\ref{2D} we provide examples for different emission sources detected in these 2D spectra. A typical H{\sc II} emitter is indicated the region marked with a red circle, while a background galaxy is marked with a red rectangle. Each spectrum found in the images is wavelength calibrated and distortion corrected, using the Gemini  package (\textit{gemini.gmos.gswavelength} and \textit{gemini.gmos.gstransform}). Then, 1D spectra are extracted and finally the velocities are measured, via a Gaussian fit, using the position of the redshifted H$\alpha$ (6563\AA{}) emission line, for all the H{\sc II} regions candidates (task \textit{rv.rvidlines}). The standard deviation estimated for the velocity of each region was estimated using Monte Carlo simulations and together with the error in the wavelength calibration we obtain a final error for the velocities of $\sim$ 40 km s$^{-1}$.

 Using the MSIS technique we detected 36 emission-line objects in the 2D images. Doing an inspection of the 1D spectra for the detected emission lines, we only considered 26 of them, based on their signal-to-noise (SNR). Emissions with SNR $\leq$ 10 were discarded. We note that emission lines of all our intergalactic region candidates are unresolved in wavelength.

\textit{Computing $\alpha$ and $\delta$ (J2000) for the emission line candidates}
 
For a given 2D spectrum we have the pre-imaging with the position of each slit of the mask. Thus, for each one of the emission line objects we measured the coordinate (x,y)$_{em}$ (in the
spectral plane), which was then transformed to (x,y)$_{im}$ coordinates in the pre-imaging. With the position (x,y)$_{im}$ we transformed the coordinate of each emission line to the corresponding world coordinate system WCS $\alpha$, $\delta$ (J2000). In Table \ref{table:basic} we list the $\alpha$ and $\delta$ for the H{\sc II} regions candidates as well as the central wavelength ($\lambda_c$), the full width half maximum (FWHM), the systemic velocity and the flux for the H$\alpha$ line. The same parameters are listed in table \ref{table:basic_dwarf} for the emitters in the satellite galaxy FGCE 0745. 

In this work we did not find extended sources with
  low surface brightness. For such spatially extended sources, the
  MSIS technique detects emissions in more than one slit position
  along the x-axis in the mask or as an extended emission along the y-axis. Such emission if present would be
  visible in the sky spectrum, which is obtained by extracting spectra
  from all the slits along the same column. None of the sky spectra
  showed emission at the redshifted wavelength of the H$\alpha$
  emission. Note that for only one source, (IG$\_$04), H$\alpha$ emissions were
  detected in several slits: this source is clearly a bright extended H{\sc
    II} region with several substructures parts and not a low surface brightness source.

\subsection{HI gas}

\citet{schiminovich95} obtained neutral hydrogen maps of the system NGC~2865 with the Very Large Array (VLA, beam 73\arcsec$\times$40\arcsec). The H{\sc I} cloud is distributed in a discontinuous ring around the galaxy NGC~2865 with a total mass of 1.2 $\pm$ 0.6 $\times$ 10$^9$ M$_\odot$. It is rotating around NGC~2865 with a circular velocity $\sim$ 250 km s$^{-1}$. The H{\sc I} ring around NGC~2865 extends for $\sim$110 kpc of diameter (north and south H{\sc I} tails) and it has a width of $\sim$45 kpc. Assuming that the gas is rotating in a circular orbit centered on the nucleus, the estimated inclination is 65$^\circ$. \citet{schiminovich95} indicated that the main body of NGC~2865 and the gas ring might have formed from the same event, and this can be explained by a major merger of nearly equal mass progenitors. \citeauthor{schiminovich95} using stellar spectroscopy and UBV photometry determined the merger  to be between 1 and 4 Gyr ago. These estimates are consistent with those found by \citet{Fort86} for low surface brightness structures in NGC~2865. If the H{\sc I} gas clumps are orbiting in a stable circular orbit, the upper limit for the age of the merger corresponds to 8 orbits of the gas around NGC~2865. \citet{schiminovich95} also detected the presence of a gas-rich satellite galaxy, FGCE 0745, located 6\arcmin\ to the southeast of NGC~2865. It is an edge-on spiral galaxy containing 4.4$\times$10$^8$ M$_\odot$ of H{\sc I}.

\subsection{Ultraviolet data}

We used GALEX FUV and NUV background-subtracted images from the MultiMission Archive at the Space Telescope Science Institute (MAST). The exposure times were 2560s and 16250s for FUV and NUV, respectively. The fluxes were measured using the task \textit{PHOT} in IRAF. The aperture used is optimised to the size of the region in the NUV image. In cases where the detection was not spatially resolved we assume the aperture as the Point Spread Function (PSF) of GALEX (5\arcsec\ FWHM), centered on the coordinates of H$\alpha$ emission. The FUV and NUV magnitudes were
calculated using \citet{Morrissey05}
m$_\lambda$=2.5log[F$_\lambda$/a$_\lambda$]+b$_\lambda$, where
a$_{NUV}$=2.06$\times$10$^{-16}$erg s$^{-1}$ cm$^{-2}$ \AA$^{-1}$,
a$_{FUV}$=1.4$\times$10$^{-15}$erg s$^{-1}$ cm$^{-2}$ \AA$^{-1}$,
b$_{NUV}$=20.08, and b$_{FUV}$=18.82, inside an aperture which
depended on the size of the source given by the H$\alpha$ line.
Fluxes are multiplied by the effective filter bandpass  ($\Delta\lambda_{FUV}$ = 269\AA\ and $\Delta\lambda_{NUV}$ = 616\AA)
to give units of erg s$^{-1}$ cm$^{-2}$. The
magnitudes in FUV and NUV were corrected for foreground Galactic
extinction using E(B$-$V)=0.0716 and A$_{FUV}$=E(B$-$V)$\times$8.29,
and A$_{NUV}$=E(B$-$V)$\times$8.18 \citep{Seibert05}.

\section{Analysis}

For each object, we derived the following parameters : 1) stellar mass, 2) ionizing photon luminosity, 3) total mass of ionized hydrogen, 4) age and 5) metallicity, when possible. In this section we explain how these quantities are computed. 

 In order to obtain the main physical parameters for each region, we derived the H$\alpha$ luminosity (L$_{H\alpha}$ = 4$\pi D^2 F_{H\alpha}$), considering a distance of 35.0 Mpc , distance estimated by \citet{Georgakakis01} for the galaxy NGC~2865 (assuming H$_o$ = 75 km s$^{-1}$ Mpc$^{-1}$). From L$_{H\alpha}$ we derived the principal parameters using the equations below: 

\textit{The Star Formation Rate} (SFR): One of the main unknowns in the study of star formation in young clusters is the value of the intrinsic dust absorption. Since there are no infrared data available for these regions and the H$\beta$ line is not observed, we were not able to estimate the internal extinction, thus only a lower limit to the total SFR is estimated. We used the equation given by \citet{Kennicutt98} to estimate the SFR for all regions, assuming a ``continuous star formation'' approximation. In such case, we use L$_{H_{\alpha}}$ as the sum of the luminosities of all sources. Considering all {\sc HII} regions together, we estimated a lower limit of the total SFR for the ensamble as follows
\begin{eqnarray}
SFR_{H_{\alpha}} \text{(M$_\odot$ yr$^{-1}$)} = \frac{L_{H_{\alpha}}}{1.26 \times 10^{41}}; \; \;  \; \;  \; \;  \; \;  \; \;  \; \;  \; \;\text{[L$_{H{\alpha}}$ in erg s$^{-1}$]}
\end{eqnarray}
With the purpose of estimating the total mass of ionized hydrogen we computed the \textit{ionizing photon luminosity}, Q(H$^\circ$), given by \citet{Osterbrock06}:
\begin{eqnarray}
Q(H^0) \text{(photon s$^{-1}$)} = 7.31\times10^{11} L_{H\alpha} ;  \; \;  \; \;  \;   \; \; \text{[L$_{H{\alpha}}$ in erg s$^{-1}$]}
\end{eqnarray}

\textit{The total mass of ionized hydrogen} (M$_{H{\sc II}}$), estimated as suggested by \citet{Osterbrock06}:
\begin{eqnarray}
M_{H{\sc II}}  \text{(M$_\odot$)}  = \frac{Q(H^0) \, m_p}{n_e \, \alpha_B}
\end{eqnarray}

Where  n$_e$ and m$_p$ denote the electron density and proton mass. We consider $n_e$ = 400 cm$^{-3}$, typical values of the electron density for brighter regions in 30 Dor \citep{Osterbrock06}, and the recombination coefficient $\alpha_B$ = 2.59 $\times$ 10$^{-13}$ cm$^{-3}$ s$^{-1}$. 

Table \ref{table:Ha} and \ref{table:Dwarf} list the physical parameters described above using the H$\alpha$ line of the intergalactic H{\sc II} region candidates and the star forming regions in the satellite galaxy FGCE 0745.

\subsection{UV emitting regions} \label{sec:age_mass}

Whenever possible, we estimated the luminosity in the FUV and NUV-bands for each H{\sc II} source (assuming that the distance to the main galaxy NGC~2865 is D = 35 Mpc). With the emission in UV and considering an instantaneous burst, we estimated a lower limit for the age and the mass for each intergalactic H{\sc II} region. We note that here we used a different assumption from the one used for the estimate of the SFRs. To estimate ages and masses, we are treating each region individually and we assume instantaneous burst. For the regions IG\_17\_P1 and IG\_51\_P3, we did not detect emission coming from the FUV-band. Therefore, for these two regions we were not able to estimate ages and masses.

\textit{Age estimates} --- For the five regions with detected UV-emission, a lower limit to their age comes from the color FUV-NUV and the models given by STARBURST99 \citep[SB99;][]{Leitherer99}. These models are generated for an instantaneous burst, solar metallicity and Salpeter initial mass function (IMF; 0.1-100 M$_\odot$), and they are optimized for GALEX filter transmission curves. We remark that we did not correct the UV-magnitudes for the internal reddening, therefore our estimates are only lower limits to the age of the star forming event. We note that \citet{Boquien07} adopt a method based on both NUV and H$\alpha$ emissions to estimate the ages for the intergalactic H{\sc II} regions. However, in our case this estimator is not applicable because the aperture used to measure the emission in ultraviolet and optical was different: the FWHM for the GALEX images is typically $\simeq$ 5\arcsec and the H$\alpha$ emission was measured from a spectrum within an aperture of 1\arcsec\.). 
\textit{Stellar mass estimates} --- We used the obtained ages and the FUV luminosities to estimate the stellar masses with SB99 models. These values were obtained from the SB99 monochromatic luminosity, L$_{1530}$, for an instantaneous burst, Salpeter IMF (from 0.1 to 100 M$_{\odot}$), and solar metallicity for stellar masses between 10$^3$ to 10$^8$ M$_\odot$. With the age obtained by the color FUV-NUV we fixed the luminosity, in 1530\AA{}, given by SB99 for the different mass and using the intrinsic luminosity of each object we interpolated the respective mass. Since we used ultraviolet fluxes for the computations, older generations of stars were not taken into consideration in the stellar mass determination.

\subsection{Metallicity}

We calculated the metallicity, when possible, using the empirical method N2, proposed and calibrated by \citet{Pettini04}. This `empirical' method is adequate for estimating oxygen abundances in extragalactic H{\sc II} regions. The method considers the N2 index (N2 $\equiv$ log$\lbrace$[N{\sc II}] $\lambda$6583/H$\alpha\rbrace$) and the relation 12+log(O/H)=8.90+0.57$\times$N2 to estimate the oxygen abundance. The uncertainties on the calibration of this method are 0.18 dex when 68$\%$ of the points are included. The same method was used in \citet{demello12} for tidal dwarf galaxies and intergalactic H{\sc II} regions, and they found a close agreement in the oxygen abundance derived from the O3N2\footnote{O3N2  $\equiv$ log$\lbrace$([OIII]$\lambda$5007/H$\beta$)/([NII]$\lambda$6583/H$\alpha)\rbrace$} and N2 method. 

\section{Results}

Using the MSIS technique, we carried out a flux limited survey of the  H$\alpha$ emitters over the southeastern H{\sc I} tail around NGC~2865, down to a flux limit of 10$^{-18}$ erg cm$^{-2}$ s$^{-1}$ \AA$^{-1}$. We consider a field of view of 5\arcmin $\times$ 5\arcmin centered at $\alpha$(J2000) 09$^{h}$ 23$^{m}$ 37\fs13 and $\delta$(J2000) -23\degr\ 11\arcmin\ 54\farcs34. We found 26 spectra with SNR $>$ 10. Seven of them are emission sources in the intergalactic medium around NGC~2865, and nineteen are regions associated with the satellite galaxy FGCE~0745.

Each source is identified with a label, either IG or Sat \_${\textit{slit}}$\_${\textit{mask}}$, depending whether such emission is found in the intergalactic medium around NGC~2865 or it is associated with the satellite galaxy. Here \textit{mask} is the number of the mask position where the source was found and \textit{slit} is the slit number for a given position of the mask. In some cases, for the same slit, we detected two emissions at adjacent offset positions: in such cases we add a number (1 or 2) next to the slit number.

   \begin{figure}[h!]
   \centering
       \includegraphics[width=\hsize]{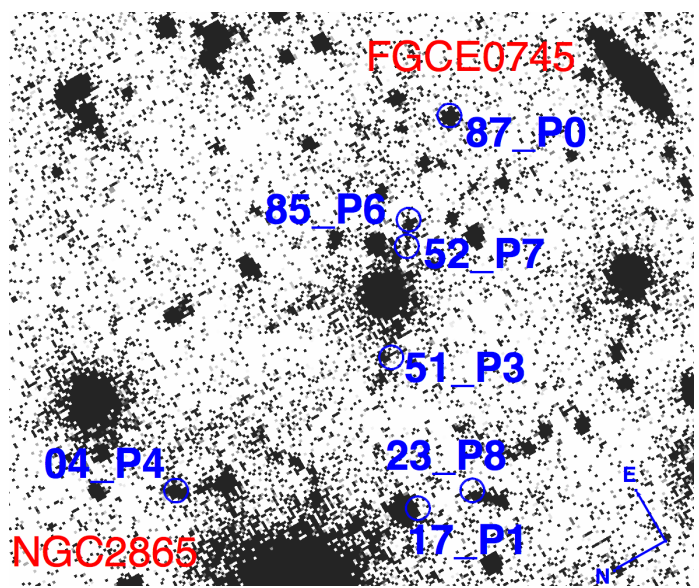}
\caption{The GALEX-NUV image in a field of view of 5.5\arcmin\ $\times$ 5.5\arcmin. The seven extragalactic H$\alpha$ sources are indicated by blue labels; red labels indicate the galaxies NGC~2865 and FGCE~0745.}
         \label{NUV_26}
   \end{figure}

\subsection{Intergalactic H{\sc II} regions}\label{sec:HII}

\begin{figure*}
   \includegraphics[width=\hsize]{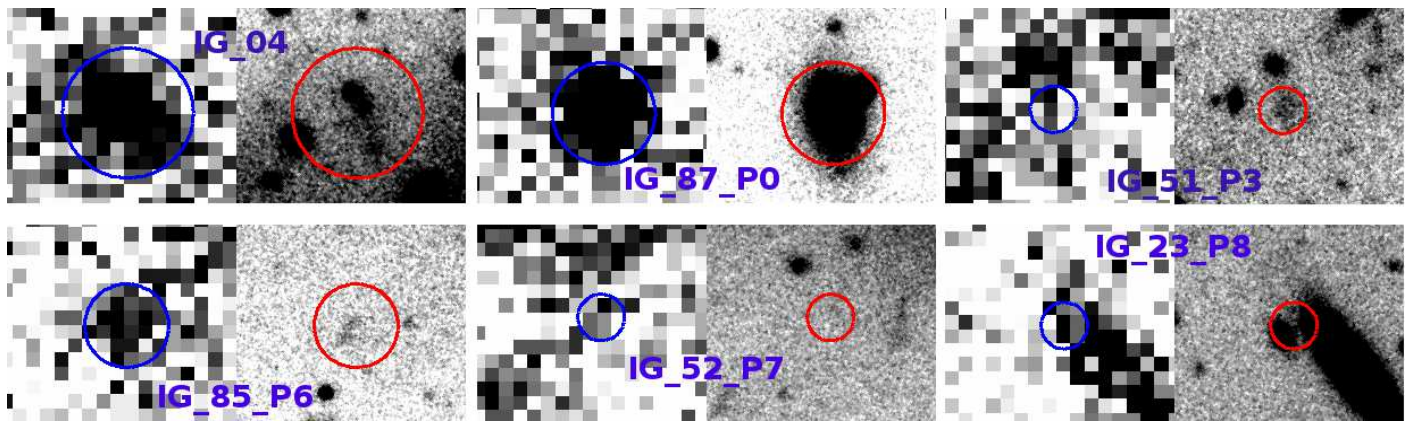}
   \caption{Images for six of the seven intergalactic regions in GALEX-NUV and Gemini-r bands, indicated by the blue and red circles, respectively. The radius of each circle indicates the aperture used in the photometry for each source. The size of each image is  21\arcsec\ $\times$ 25\arcsec.}
   \label{NUV_r}
   \end{figure*}

\begin{figure}[h!]
   \centering
 \includegraphics[width=\hsize]{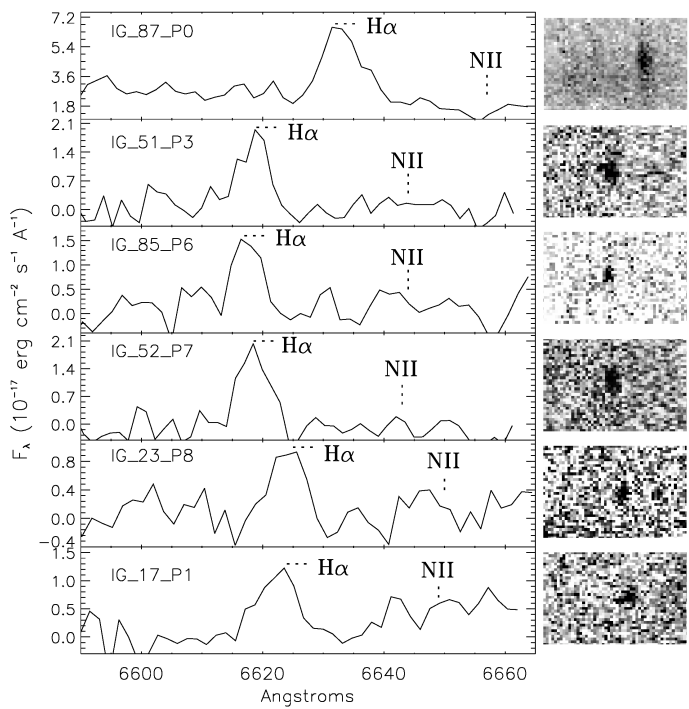}
      \caption{Emission lines from detected intergalactic H{\sc II} region candidates: 1-D (left) and 2-D (right). The H$\alpha$ line and the place where it is expect to find the [N{\sc II}] line are marked.}
         \label{HII_region1}
   \end{figure}

\begin{figure}[h!]
   \centering
 \includegraphics[width=\hsize]{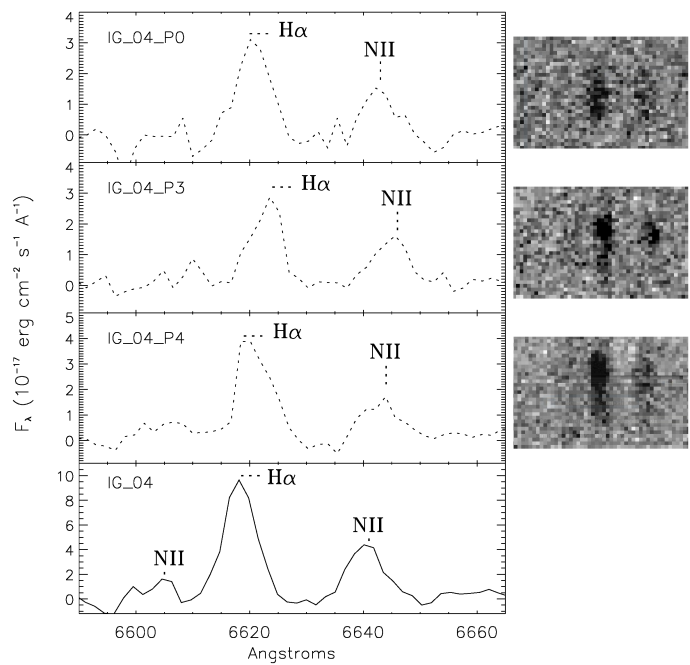}
      \caption{Emission lines of the region IG$\_$04: 1-D (left) and 2-D (right). The position of the H$\alpha$ and [N{\sc II}] lines are marked.}
         \label{HII_region1_04}
   \end{figure}

Seven sources were found in the intergalactic region around NGC~2865. In Figure~\ref{NUV_26} we plot the seven intergalactic H$\alpha$ emissions on the NUV image. In Figure~\ref{NUV_r} we show the GALEX NUV and \sloanr-Gemini counteparts at the position of the detected H$\alpha$ emission; blue circle indicates the GALEX NUV and the red circle the \sloanr-Gemini; circles have of the same sizes as the apertures used in the photometry. The size of each image is 0.35\arcmin $\times$ 0.41\arcmin. The 2D and the extracted spectra for six of the seven regions are plotted in Figure~\ref{HII_region1}. The H$\alpha$ line is marked in each case. We also mark the position where the emission line [N{\sc II}]$\lambda$6583 it is expected. For one of the seven regions, IG$\_$04, we detect emission at three different positions of the mask (POS0, POS3 and POS4). In Figure~\ref{HII_region1_04} we show these three spectra with dashed lines and their sum with a solid line. The H$\alpha$ and [N{\sc II}] lines are also marked.

We note that in this paper we rule out the possibility that these regions are background Ly$\alpha$ emitters. We compared the flux obtained in each detected line with the flux estimate for a Ly$\alpha$ emitter at $z$ = 4 (considering that the Ly$\alpha$ line is at $\lambda$1215\AA\ and they are observed at $\lambda$6620\AA). We used the result from \citet{Gronwall07} for a survey of Ly$\alpha$ at $z$ = 3.1, where the mean flux obtained for this sample was 2.9$\times$10$^{-17}$ erg s$^{-1}$ cm$^{-2}$ . Using the cosmology calculator \citep{Wright06} we extrapolated the mean flux from \citeauthor{Gronwall07} for emitters at $z$ = 4. Their estimated flux is then 1.5$\times$10$^{-17}$erg s$^{-1}$ cm$^{-2}$, while the flux of our detected sources is an order of magnitude brighter, i.e. a mean flux of 1.8$\times$10$^{-16}$erg s$^{-1}$ cm$^{-2}$. Given this argument, we then conclude that all emissions detected in our work are from the H$\alpha$ line.

We now describe each intergalactic region in turn: 

\textit{Region IG\_04} --- In the Gemini \sloanr-band, Figure~\ref{NUV_r}, the region is resolved into subcomponents: a stellar cluster and a tail. The size of the region, in the Gemini image, was 3\farcs9 (714 pc), without considering the tail. This region appears to be similar to the two regions found in HGC100 by \citet{demello12}. The H$\alpha$ emission from region IG\_04 was observed at 3 different positions of the mask, POS0, POS3, and POS4 (\textit{IG\_04\_P0, IG\_04\_P3 and IG\_04\_P4}). At each position, the H$\alpha$ line appears with good SNR $\simeq$ 30. When we combined the 1D spectra using the task \textit{scombine} in IRAF and with `sum' option in the combine operation, it is possible to detect and measure the fluxes of the [N{\sc II}] line ($\lambda$ = 6583\AA{}, SNR$\simeq$15) and the [N{\sc II}] line ($\lambda$ = 6548\AA{},  SNR$\sim$10). The detection of these three lines confirm that IG\_04 is an intragroup H{\sc II} region, and not a background object.  Despite the low SNR of the [N{\sc II}]$\lambda6548$ line, we measured the relative fluxes between [N{\sc II}] lines.  We estimated that the [N{\sc II}]$\lambda6548$/[N{\sc II}]$\lambda6583$ ratio is $\simeq$3.3, which is within the errors, if we consider the predicted model ration of 3 \citep{Osterbrock06}. Thus, we are able to estimate the oxygen abundance.The N2 index is -0.3, which are 12 + log(O/H)=8.7, indicating solar metallicity \citep{Asplund09}. We highlight that for the region IG\_04 we found two different velocities for the three mask positions. For the positions POS0 and POS4 we measured a velocity of $\sim$2630 $\pm$ 40 km s$^{-1}$,  but for the position POS3 the velocity was 2811 $\pm$ 40 km s$^{-1}$, different by $\sim$200 km s$^{-1}$. This fact may suggest that we are observing two different sources.

\textit{Region IG\_87\_P0} --- For the UV and \sloanr-band images we used an aperture of radius 5\farcs5 to obtain the principal physical parameters of this region. In the H$\alpha$ line this region has the highest SNR of the seven intergalactic H{\sc II} regions, SNR $\simeq$ 48. The strong emission in H$\alpha$, optical and UV-bands can imply that two different scenarios could be happening: 1) it could be an early-type galaxy at high redshift emitting a strong emission line which is not H$\alpha$; 2) the region has three different stellar populations: one contains older stars (optical emission), a new generation of stars which emitt in the UV, with age of approximately 200 Myr (estimated in section \ref{sec:age_mass}), together with a very young population, $\sim$ 10Myr, which emits in the H$\alpha$ line. We note that this object is at least 2 mag brighter than any of the other six regions studied here and its morphology resemble that of a galaxy, which would indicate that it is not an H{\sc II} region-like object. However, a spectrum covering a greater range of wavelengths is necessary to confirm the nature of this region. 

\textit{Region IG\_51\_P3} ---  This object is not detected in FUV while in NUV there is a faint emission which blended with an extended source in the North. In the \sloanr-image the emission is spatially resolved and appears like a small cluster. One possible explanation for the absence of the FUV emission is a possibly large amount of dust present in the region. Since the source is not resolved, we considered a radius of 2\farcs5 , given by the resolution of GALEX (FWHM $\simeq$ 5\arcsec). 

\textit{Region IG\_85\_P6} --- The emission line, H$\alpha$, has a SNR$\simeq$ 13. This source is resolved in both FUV and NUV-bands, and presents a faint extended emission in the optical \sloanr-band image. For the photometry we used a radius of 4\farcs5, given the size of the source in the UV images. We estimated a lower limit to the age of about 16  Myr and a mass of 9$\times$10$^{4}$M$_\odot$, typical of a young cluster.   

\textit{Region IG\_52\_P7} --- This region presents a faint extended emission in UV bands and a very diffuse emission in the \sloanr-band. Although it is an unresolved source, we estimated the flux from the UV in a radius of 2\farcs5 (given by the resolution of GALEX) centered on the position of the H$\alpha$ line. 

\textit{Region IG\_23\_P8} ---  In GALEX images the UV-emission is blended with an extended galaxy emission near to the source. In \sloanr-band it is possible to resolve two sources, a galaxy and a nearby small emission, which could be the responsible for the detected H$\alpha$ line. For the photometry of this unresolved region we used a radius of 2\farcs5, centered on the H$\alpha$ emission.   

\textit{Region IG\_17\_P1} --- The spectrum of this region contains a weak H$\alpha$ line with a SNR $\sim$ 11. It is not observed either in FUV nor in the  \sloanr-band emission. In the NUV and FUV images one can see that the source clearly overlaps the position of a bright foreground star (see Figure~\ref{NUV_26} for the NUV image), which prevents us from classifying this region as a young and low-mass region.

All these regions are located in low-density of H{\sc I} gas \citep[HI $\geq$ 10$^{19}$ cm$^{-2}$,][]{schiminovich95}, where in general the probability of forming stars is low \citep{Maybhate07}.

When we compare the ages derived for the several regions above, we note that they can be as large as 50 to 200 Myr, while our selection was based on H$\alpha$, which is known to be present in systems with ages smaller than about 10 Myr. This could seem like a contradiction but, in fact, it is not. The first argument is that the FUV-NUV flux comes from an extended 5\arcs region which is larger than the region emitting in H$\alpha$, which is unresolved on the MSIS images. The second argument is that extinction may play a role and given that we only have one color, we cannot correct for extinction. The third and last argument is that the FUV-NUV probes in fact an older population than the one causing emission in H$\alpha$, i.e. we detect the FUV, NUV continuum from stars that have an age $<$ 200 Myr (from Starburst99) which also containts a subpopulation that is only a few Myr old and emits in H$\alpha$. In other words, what we are seeing in the UV is a combination of young and massive stars (also seeing in H$\alpha$) and more evolved stars formed earlier. We note that the UV luminosities found for these HII regions are larger than those expected from the recent star forming population giving rise to the measured H$\alpha$ flux. From Figure 9 in \cite{Pflamm09}, we estimate the expected FUV luminosity associated with the star forming episode associated with the measured H$\alpha$ luminosity of 5 $\times$ 10$^{37}$ erg s$^{-1}$. The expected FUV luminosity is 2 $\times$ 10$^{25}$ erg s$^{-1}$, nearly 10 times lower than the measured FUV fluxes for these sources. 
 
\begin{figure}
   \centering
      \includegraphics[width=\hsize]{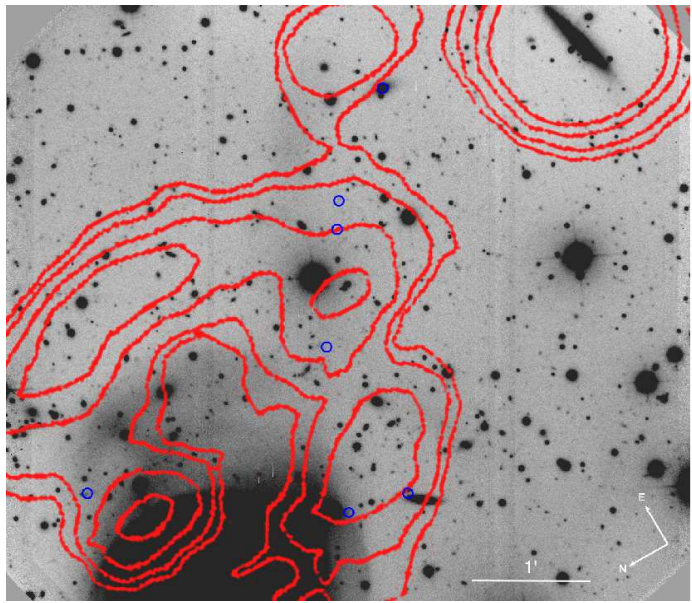}
\caption{H{\sc I} contours from \citet{schiminovich95}.The gas is shown as red contours on an optical \sloanr-band from Gemini.The H{\sc II} regions candidates are indicated by the blue circles. The Gemini image only considers the south tidal debris of the system. This Figure shows that the H{\sc II} sources are detected along the ring of H{\sc I} contours. The contour levels are 1.9, 3.8, 7.6 and 11.4 $\times$ 10$^{19}$ cm$^{-2}$. The VLA beam is 73\arcsec $\times$ 40\arcsec.}
         \label{contours}
   \end{figure}

In Figure~\ref{contours} we show the intergalactic H$\alpha$ emitter regions, found in this work, superposed onto the H{\sc I} contours and Gemini \sloanr-band \citep[HI contours taken from][]{schiminovich95}.

\subsection{Are the intergalactic H{\sc II} regions bound to NGC~2865?}

\begin{figure}
   \centering
   \includegraphics[width=\hsize]{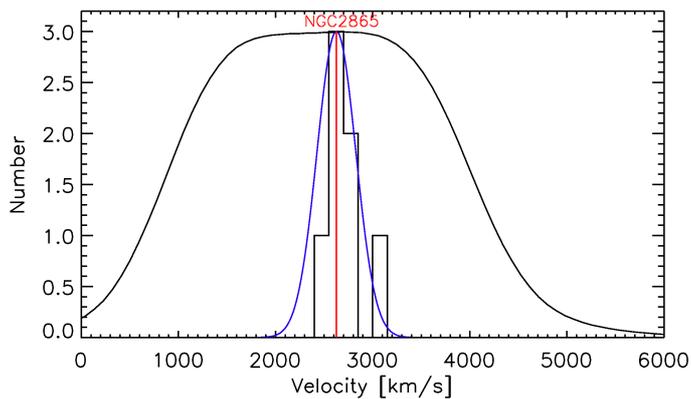}
      \caption{The histogram for the line-of-sight velocities $v_{LOS}$ of the H$\alpha$ emitters together with the Ha\_C filter used for the MSIS observations. The red line indicates the systemic velocity of NGC~2865 (2627 km s$^{-1}$, \citet{Smith00}). The distribution of the velocity of the seven intergalactic H{\sc II} regions has a peak at 2711 km s$^{-1}$, very close to the systematic velocity of NGC~2865. The Gaussian distribution is plotted in blue has average velocity 2627 km s$^{-1}$ and $\sigma$=200 km s$^{-1}$.} 
               \label{Histogram}
   \end{figure}

\begin{figure} \centering
      \includegraphics[width=\hsize]{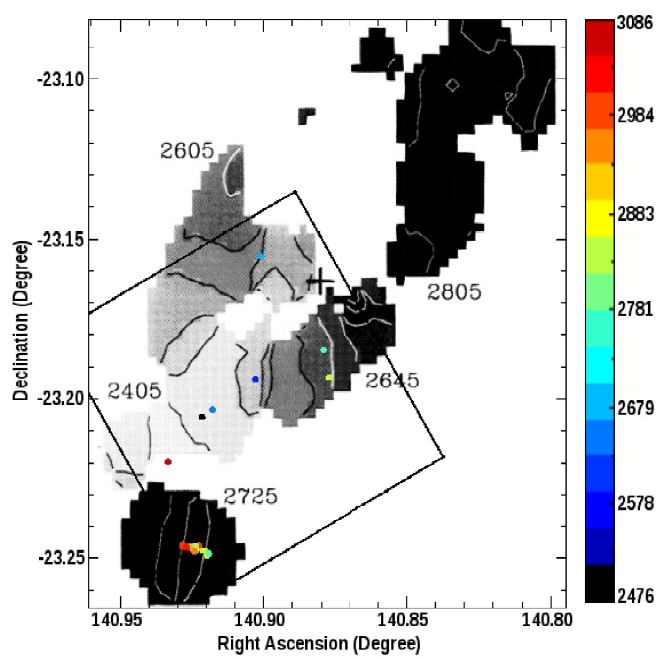} \caption{
        The LOS velocities of the H{\sc II} regions discovered in the
        current survey and H{\sc I} velocity field taken from
        Schiminovic et al. (1995), figure 1d. The cross at the center of the ring
        indicate the position of the galaxy NGC~2865. The color code
        refers to the LOS velocity. The velocity gradient along the
        major axis of FGCE 0745 is clearly visible and ranges from
        2781 km s$^{-1}$ to 3063 km s$^{-1}$. The field of view
        surveyed with the MSIS technique is 5\arcmin\ $\times$5\arcmin
        and is shown by the black rectangle.}
	 \label{overplot}
   \end{figure}

We now discuss the observational evidence in support of these systems to be gravitationally bound to the central galaxy NGC~2865. Using the H$\alpha$ emission line and the task noao.onedspec.rv.rvidlines within IRAF we measured the line-of-sight (LOS) velocities for each H{\sc II} region. In Figure~\ref{Histogram} we plot the histogram of their V$_{LOS}$ together with the filter Ha\_C bandpass used in the observation. The systemic velocity of NGC~2865 \citep[2627 km s$^{-1}$,][]{Smith00} is indicated with a red line and it is very close to the peak of the distribution of the LOS velocities of the intergalactic H{\sc II} regions. In fact, the average velocity of the seven H{\sc II} regions is 2711 km s$^{-1}$, which is only 84 km s$^{-1}$ redder than the systemic velocity of NGC~2865. Our seven intergalactic H{\sc II} regions have also LOS velocities that are similar to those of the H{\sc I} gas at the same location (velocity of the gas from 2405 to 2645 km s$^{-1}$), hence this supports an association between these H{\sc II} emitters and the H{\sc I} tail. In Figure~\ref{overplot}, the H$\alpha$ sources are over plotted on the velocity field of the H{\sc I} gas. The velocity range in the H{\sc I} tail is also consistent with the Gaussian distribution with average velocity = 2627 km s$^{-1}$ and $\sigma$ = 200 km s$^{-1}$: hence the kinematic of the H{\sc I} tail is consistent with that of on orbit gravitationally bound to the central galaxy NGC~2865.

If the seven H{\sc II} regions are at equilibrium, the velocity dispersion obtained from the standard deviation of their LOS velocities is 184 km s$^{-1}$. This value is similar to the central velocity dispersion measured by \citet{Hau99} for NGC~2865, $\sigma_0= 200$ km s$^{-1}$. In Figure~\ref{Histogram} we plot a Gaussian distibution with $\sigma = 200$  km s$^{-1}$ and average 
velocity $\bar{V}= 2627$km s$^{-1}$, which turns out to be a good approximation for the histogram distribution of the LOS of the seven H{\sc II} regions. 

The result of the comparison between the histogram of the H{\sc II} $v_{LOS}$ and the Gaussian distribution is consistent with the hypothesis that these intergalactic H{\sc II} regions are bound to NGC~2865.

\begin{table*}
\caption{Intergalactic H{\sc II} regions candidates in the environment of NGC~2865.}             
\label{table:basic}      
\centering          
\begin{tabular}{l c c c c c c}  
\hline\hline       

\multicolumn{1}{c}{ID} & $\alpha$ & $\delta$ & $\lambda$ & FWHM  &$V_{sys}$\tablefootmark{a}  & Flux$_{H\alpha}$ \\
 & (J2000) & (J2000) & \AA{} & \AA{}  & km s$^{-1}$ & erg s$^{-1}$ cm$^{-2}$  \\
\hline                    
 NGC 2865\tablefootmark{b} & 09$^{h}$ 23$^{m}$ 30\fs2  & -23\degr\ 09\arcmin\ 41\farcs0  & ---  &   & 2627   &  ---	   \\
 IG\_04\_P0 & 09$^{h}$ 23$^{m}$ 36\fs3   & -23\degr\ 09\arcmin\ 15\farcs8   & 6621  & 7.2  & 2666 	& 2.6e-16  \\
 IG\_04\_P3 & 09$^{h}$ 23$^{m}$ 36\fs3   & -23\degr\ 09\arcmin\ 15\farcs8   & 6624  & 5.9  & 2811	& 1.9e-16  \\
 IG\_04\_P4 & 09$^{h}$ 23$^{m}$ 36\fs3   & -23\degr\ 09\arcmin\ 15\farcs8   & 6620  & 7.2  & 2628	& 3.2e-16  \\
 IG\_87\_P0 & 09$^{h}$ 23$^{m}$ 43\fs9   & -23\degr\ 13\arcmin\ 08\farcs9   & 6630  & 7.3  & 3070	& 3.6e-16  \\
 IG\_17\_P1 & 09$^{h}$ 23$^{m}$ 30\fs9   & -23\degr\ 11\arcmin\ 05\farcs6   & 6623  & 7.9  & 2789	& 1.1e-16  \\
 IG\_51\_P3 & 09$^{h}$ 23$^{m}$ 36\fs6  &  -23\degr\ 11\arcmin\ 38\farcs2  & 6619  & 6.3  & 2606	& 1.5e-16  \\
 IG\_85\_P6 & 09$^{h}$ 23$^{m}$ 41\fs1  &  -23\degr\ 12\arcmin\ 20\farcs8   & 6617  & 5.1  & 2507	& 1.0e-16  \\
 IG\_52\_P7 & 09$^{h}$ 23$^{m}$ 40\fs2  &  -23\degr\ 12\arcmin\ 12\farcs7   & 6618  & 5.8  & 2568	& 1.4e-16  \\
 IG\_23\_P8 & 09$^{h}$ 23$^{m}$ 30\fs5  &  -23\degr\ 11\arcmin\ 36\farcs4   & 6624  & 5.7  & 2828	& 9.0e-17  \\
\hline                  
\end{tabular}
\tablefoot{
\tablefoottext{a}{Systemic velocities deduced from our H$\alpha$ line analysis. Heliocentric velocities given by the task rvidline from IRAF. The errors for the velocities were estimated using monte carlo simulation, for 100 runs, and they are about 40 km s$^{-1}$ for all the spectra.}
\tablefoottext{b}{For the main galaxy, NGC~2865, the value for its radial velocity was taken from \citet{Smith00}.}
}
\end{table*}

\subsection{Star Forming regions in the satellite galaxy FGCE~0745}\label{sec:Dwarf} 

FGCE~0745 is an edge-on galaxy located $\simeq$6\arcmin\ to the southeast of NGC~2865 (see Figure~\ref{NUV_26}). As shown in Figure~\ref{contours}, the H{\sc I} gas distribution of this galaxy is relatively undisturbed, with no evidence of on-going interaction with NGC~2865. This does not exclude that these two galaxies have not interacted in the past, given that the velocity of the H{\sc I} gas of FGCE~0745 is similar to the velocity of the H{\sc I} ring around NGC~2865. 

   \begin{figure}
   \centering
   \includegraphics[width=\hsize]{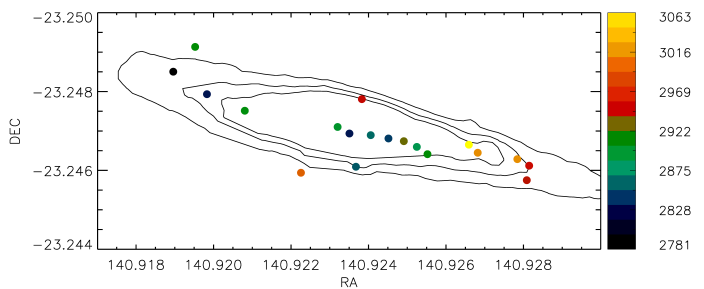}
      \caption{Velocity field for the satellite galaxy~FGCE 0745. The dots are the H$\alpha$ detections in the galaxy and the different colors represent different velocities (km s$^{-1}$) listed on the bar on the right side of the figure. The contour levels are 22.3, 22.2 and 22.1 mag/arcsec$^{2}$ from the Gemini \sloanr-band.}
         \label{velocity_map}
   \end{figure}

\begin{figure}
   \centering
   \includegraphics[width=\hsize]{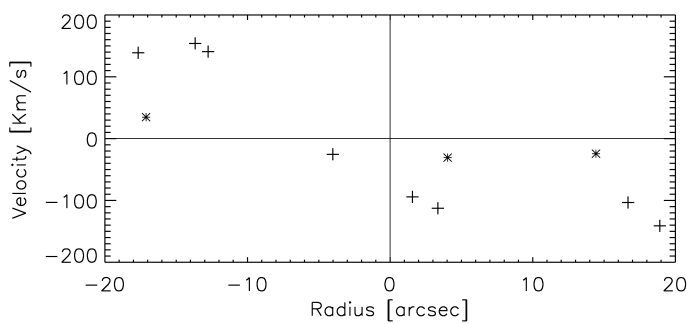}
      \caption{Rotation curve for the satellite galaxy FGCE~0745. Excluding the three measurements indicated by asterisk, the rotation curve is very similar to those of late-type galaxies.  We infer a V$_{max}$ $\simeq 160 $km s$^{-1}$.}
         \label{rot_curve}
   \end{figure}

Using the MSIS technique we found ninenteen H$\alpha$ sources within the satellite galaxy FGCE 0745. The spectra of these ninenteen sources are shown in Figure~\ref{spectra_dwarf21}. The coordinates, velocity and H$\alpha$ fluxes for each source are listed in Table~\ref{table:basic_dwarf}. The optical emission of the galaxy is at the same location as  the peaks in HI gas and the H$\alpha$ emission shown in Figure~\ref{contours} and \ref{velocity_map}. Using the LOS velocity of each detected H$\alpha$ source we derive a velocity map for FGCE~0745. In Figure~\ref{velocity_map} this map is shown and the contours represent the optical emission of the galaxy, taken from the  \sloanr-band. We note that most of the H$\alpha$ sources were detected along the photometric major axis of the galaxy. Assuming circular orbits (considering the angular inclination as $\cos^{-1}(\frac{b}{a})$), we determine the rotation curve (Fig.~\ref{rot_curve}) using an inclination of 80\degr, a position angle of 70\degr, the systemic velocity 2920 km s$^{-1}$ and the kinematic center of FGCE~0745 to be at $\alpha$(J2000) 9$^{h}$ 23$^{m}$ 41\fs85 $\delta$(J2000) -23\degr\ 14\arcmin\ 47\farcs97. The rotation curve is derived taking into account those pixels within a cone of 30\degr\ aligned with the galaxy's major axis. Excluding the three point in asterisk, which do not follow the circular pattern defined by the remaining points, Fig.~\ref{rot_curve} shows a rotation curve which is consistent with those of late type galaxies, and a total $\Delta V_{rot} = 320$ kms$^{-1}$. Under the assumption that the H$\alpha$ emission is detected all over the galaxy disk, we estimated V$_{max} = 160$ kms$^{-1}$. A radius of 26\arcsec\ ($\simeq$ 4.2 kpc) was derived from the optical image (\sloanr-band). A similar value for the radius of this galaxy is given in NED\footnote{Without reference}. Assuming a spherical mass distribution for this galaxy, we estimated a total mass of M$_{dyn}$ = 2.5$\times$10$^{10}$ M$_\odot$, considering the gravitational constant given by G=4.3$\times$10$^{-9}$ km$^2$s$^{-2}$ Mpc M$^{-1}_\odot$ \citep{Mo10}.

\begin{figure}
   \centering
   \includegraphics[width=\hsize]{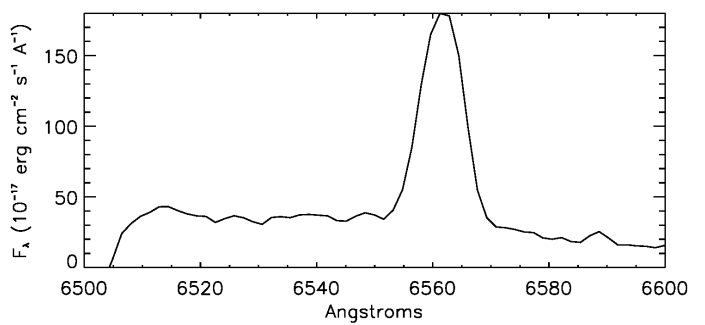}
      \caption{Coadded spectrum for the nineteen H$\alpha$ emissions detected in FGCE~0745.}
         \label{spectra_dwarf}
   \end{figure}

In Figure~\ref{spectra_dwarf} we show the coadded spectrum from the nineteen H$\alpha$ detections, shifted at the rest wavelength. The H$\alpha$ and the [N{\sc II}]$\lambda$6583 lines are observed, although the fainter line of the [N{\sc II}] doublet, at 6548\AA{}, is not detected. Using the empirical method N2, we estimated a metallicity of 12+log(O/H) = 8.0.

\begin{table*}
\caption{H$\alpha$ emissions from the satellite galaxy FGCE~0745}             
\label{table:basic_dwarf}      
\centering          
\begin{tabular}{l c c c c c c}     
\hline\hline       

\multicolumn{1}{c}{ID} & $\alpha$ & $\delta$ & $\lambda$ & FWHM  &$V_{sys}$\tablefootmark{a}  & Flux$_{H\alpha}$ \\
 & (J2000) & (J2000) & \AA{} & \AA{}  & km s$^{-1}$ & erg s$^{-1}$ cm$^{-2}$  \\
\hline         
 FGCE 0745\tablefootmark{b} & 09$^{h}$ 23$^{m}$ 40\fs8  	&   -23\degr\ 14\arcmin\ 46\farcs0  & --- &   & 2480  &  ---	   \\
 Sat\_97\_P0    & 09$^{h}$ 23$^{m}$ 41\fs6    &	-23\degr\ 14\arcmin\ 49\farcs6	& 6625  & 8.0  &   2890	&	2.6e-15	\\	
 Sat\_96\_P1    & 09$^{h}$ 23$^{m}$ 42\fs7   &	-23\degr\ 14\arcmin\ 46\farcs6	& 6627  & 7.5  &   3008	&	9.3e-16 \\
 Sat\_97\_P1    & 09$^{h}$ 23$^{m}$ 41\fs6   &	-23\degr\ 14\arcmin\ 49\farcs0	& 6624  & 7.9  &   2825	&	2.2e-15	\\
 Sat\_98\_P1    & 09$^{h}$ 23$^{m}$ 40\fs7   &	-23\degr\ 14\arcmin\ 52\farcs6	& 6624  & 7.6  &   2825	&	9.2e-16	\\
 Sat\_96\_P2    & 09$^{h}$ 23$^{m}$ 42\fs7   &	-23\degr\ 14\arcmin\ 46\farcs0	& 6626  & 7.8  &   2945	&	2.5e-16	\\
 Sat\_97.1\_P2  & 09$^{h}$ 23$^{m}$ 41\fs8   &	-23\degr\ 14\arcmin\ 48\farcs8	& 6624  & 7.5  &   2854	&	7.9e-16	\\
 Sat\_97.2\_P2  & 09$^{h}$ 23$^{m}$ 41\fs3   &	-23\degr\ 14\arcmin\ 45\farcs4	& 6627  & 7.3  &   2986	&	5.1e-16	\\
 Sat\_96\_P3    & 09$^{h}$ 23$^{m}$ 42\fs7   &	-23\degr\ 14\arcmin\ 44\farcs7	& 6626  & 7.9  &   2937	&	3.0e-16	\\
 Sat\_97\_P3    & 09$^{h}$ 23$^{m}$ 41\fs9   &	-23\degr\ 14\arcmin\ 48\farcs5	& 6624  & 7.6  &   2838	&	5.3e-16	\\
 Sat\_97.1\_P4  & 09$^{h}$ 23$^{m}$ 42\fs0   &	-23\degr\ 14\arcmin\ 48\farcs3	& 6626  & 8.4  &   2920	&	8.5e-16	\\
 Sat\_97.2\_P4  & 09$^{h}$ 23$^{m}$ 41\fs7   &	-23\degr\ 14\arcmin\ 45\farcs9	& 6625  & 7.3  &   2848	&	5.2e-16	\\
 Sat\_98\_P4    & 09$^{h}$ 23$^{m}$ 41\fs0   &	-23\degr\ 14\arcmin\ 51\farcs0	& 6626  & 7.4  &   2900	&	2.4e-16	\\
 Sat\_97\_P5    & 09$^{h}$ 23$^{m}$ 42\fs0   &	-23\degr\ 14\arcmin\ 47\farcs7	& 6625  & 7.9  &   2877	&	1.5e-15	\\
 Sat\_97\_P6    & 09$^{h}$ 23$^{m}$ 42\fs1   &	-23\degr\ 14\arcmin\ 47\farcs1	& 6626  & 7.7  &   2904	&	1.0e-15	\\
 Sat\_99\_P6    & 09$^{h}$ 23$^{m}$ 40\fs7   &	-23\degr\ 14\arcmin\ 56\farcs9	& 6626  & 6.9  &   2904	&	4.1e-16	\\
 Sat\_97\_P7    & 09$^{h}$ 23$^{m}$ 42\fs4   &	-23\degr\ 14\arcmin\ 47\farcs9	& 6629  & 9.2  &   3063	&	5.4e-16	\\
 Sat\_99\_P7    & 09$^{h}$ 23$^{m}$ 40\fs5   &	-23\degr\ 14\arcmin\ 54\farcs6	& 6623  & 7.3  &   2781	&	9.4e-16	\\
 Sat\_97\_P8    & 09$^{h}$ 23$^{m}$ 42\fs4   &	-23\degr\ 14\arcmin\ 47\farcs2	& 6628  & 7.9  &   3007	&	9.5e-16	\\
 Sat\_98\_P8    & 09$^{h}$ 23$^{m}$ 41\fs7   &	-23\degr\ 14\arcmin\ 52\farcs1	& 6627  & 6.6  &   2939	&	2.4e-15	\\
 FGCE 0745\_Total & 09$^{h}$ 23$^{m}$ 40\fs8  &  -23\degr\ 14\arcmin\ 46\farcs0   & ---   &  --- & 2900 & 1.77e-14 \\  
\hline                  
\end{tabular}
\tablefoot{
\tablefoottext{a}{Systemic velocities deduced from our H$\alpha$ line analysis. Heliocentric velocities given by the task rvidline from IRAF. The errors for the velocities were estimated using monte carlo simulation, for 100 runs, and they are about 40 km s$^{-1}$ for all the spectra.}
\tablefoottext{b}{FGCE0745, satellite galaxy of NGC~2865. The radial velocity for this galaxy was taken from NED.}
}
\end{table*}

\begin{table*}
\caption{Physical parameters derived from the H$\alpha$ line, for the intergalactic H{\sc II} regions.}             
\label{table:Ha}      
\centering          
\begin{tabular}{l c c c c c}     
\hline\hline       

\multicolumn{1}{c}{ID} & L$_{H\alpha}$ & Q$_{Ho}$ & M$_{H{\sc II}}$ & M$_{stellar}$/M$_{H{\sc II}}$ & V$_{sys}$\tablefootmark{a} \\
 & [erg s$^{-1}$] & [M$_\odot$ yr$^{-1}$] & [M$_\odot$] & &  [km s$^{-1}$] \\ 
\hline                    
IG\_04\_P0 & 4.4e+37    & 3.23e+49 & 2.62e+02  & 4.58e+03 & 2666   \\
IG\_04\_P3 & 3.2e+37    & 2.31e+49 & 1.87e+02  & 6.42e+03 & 2844   \\
IG\_04\_P4 & 5.4e+37    & 3.96e+49 & 3.21e+02  & 3.74e+03 & 2695   \\
IG\_04 & 1.8e+38        & 9.53e+49 & 7.73e+02  & 1.55e+03 & ---   \\
IG\_87\_P0 & 4.4e+37    & 4.51e+49 & 3.66e+02  & 9.26e+04 & 3038   \\
IG\_17\_P1 & 1.8e+37    & 1.31e+49 & 1.06e+02  &  ---  &  2757     \\
IG\_51\_P3 & 2.6e+37    & 1.86e+49 & 1.51e+02  &  ---   &  2579    \\
IG\_85\_P6 & 1.7e+37    & 1.24e+49 & 1.01e+02  & 1.21e+03 & 2476   \\
IG\_52\_P7 & 2.4e+37    & 1.76e+49 & 1.43e+02  & 2.80e+01 & 2631    \\
IG\_23\_P8 & 1.5e+37    & 8.48e+48 & 6.88e+01  & 8.01e+04 & 2831   \\ 
\hline                  
\end{tabular}
\tablefoot{
\tablefoottext{a}{Systemic velocity deduced from our H$\alpha$ line analysis.}
}
\end{table*}

\begin{table*}
\caption{Physical parameters derived from the H$\alpha$ line, for the satellite galaxy FGCE 0745.}             
\label{table:Dwarf}      
\centering          
\begin{tabular}{l c c c c c}     
\hline\hline       

\multicolumn{1}{c}{ID} & L$_{H\alpha}$  & Q$_{Ho}$ & M$_{H{\sc II}}$  & V$_{sys}$\tablefootmark{a} \\
 & [erg s$^{-1}$] & [M$_\odot$ yr$^{-1}$] &  [M$_\odot$] & [km s$^{-1}$] \\ 
\hline                    
Sat\_97\_P0 & 4.4e+38    & 3.2e+50 & 2.6e+03 &  2890   \\
Sat\_96\_P1 & 1.6e+38    & 1.2e+50 & 9.7e+02  &  3008   \\
Sat\_97\_P1 & 3.7e+38    & 2.7+50 & 2.2e+03  &  2823   \\
Sat\_98\_P1 & 1.6e+38    & 1.2e+50 & 9.7e+02  &  2794  \\
Sat\_96\_P2 & 4.3e+37    & 3.1e+49 & 2.5e+02  &  2945   \\
Sat\_97.1\_P2 & 1.3e+38  & 9.5e+49 & 7.7e+02  &   2854      \\
Sat\_97.2\_P2 & 8.7e+37  & 6.4e+49 & 5.2e+02  &   2984  \\
Sat\_98\_P2 & 1.6e+37    & 1.2e+49 & 9.7e+01  &   2747 \\
Sat\_96\_P3 & 5.1e+37    & 3.7e+49 & 3.0e+02  &  2983  \\
Sat\_97\_P3 & 9.0e+37    & 6.6e+49 & 5.3e+02  &  2867   \\
Sat\_96\_P4 & 1.7e+37    & 1.2e+49 & 9.7e+01  &  2988   \\ 
Sat\_97.1\_P4 & 1.4e+38  & 1.0e+50 & 8.1e+02  &  2962   \\
Sat\_97.2\_P4 & 8.9e+37  & 6.5e+49 & 5.3e+02  &  2796   \\
Sat\_98\_P4 & 4.0e+37    & 2.9e+49 & 2.3e+02  &  2837   \\
Sat\_97\_P5 & 2.5e+38    & 1.8e+50 & 1.4e+03  &  2897   \\
Sat\_97\_P6 & 1.7e+38    & 1.2e+50 & 9.7e+02  &  2903      \\
Sat\_99\_P6 & 7.0e+37    & 5.1e+49 & 4.1e+02  &  2845   \\
Sat\_97\_P7 & 9.2e+37    & 6.7e+49 & 5.4e+02  &  3086    \\
Sat\_99\_P7 & 1.6e+38    & 1.2e+50 & 9.7e+02  &  2759  \\
Sat\_97\_P8 & 1.6e+38    & 1.2e+50 & 9.7e+02  &  3007   \\
Sat\_98\_P8 & 4.1e+38    & 3.0+50 & 2.4e+03  &   2939  \\
FGCE 0745 & 2.9e+39    & 2.1e+51 & 1.7e+04 &  2900 \\
\hline                                
\end{tabular}
\tablefoot{
\tablefoottext{a}{Systemic velocity deduced from our H$\alpha$ line analysis.}
}
\end{table*}

\begin{table*}
\caption{Physical parameters derived from GALEX FUV and NUV images.}             
\label{table:Galex}      
\centering          
\begin{tabular}{l c c c c c c}     
\hline\hline       

\multicolumn{1}{c}{ID} &  log(L$_{FUV}$)  & log(L$_{NUV}$) & FUV-NUV & Age\tablefootmark{a} & Mass$_{stellar}$ (FUV) & Mass$_{stellar}$ (NUV)  \\
    & [erg s$^{-1}$ \AA$^{-1}$] & [erg s$^{-1}$ \AA$^{-1}$]&  & [Myr] & 10$^6$ [M$_{\odot}$] & 10$^6$ [M$_{\odot}$]  \\
\hline                    
 IG\_04 &  37.41  & 37.11 &  0.06 $\pm$ 0.09  & 49.34$^{+33}_{-27}$   & 1.20$^{+1.29}_{-0.73}$ &  1.05$^{+1.19}_{-0.67}$  \\
 IG\_87\_P0 &  37.66  & 37.46 &  0.31 $\pm$ 0.08  & 197.28$^{+21}_{-81}$  & 17.02$^{+7.36}_{-10.28}$ &15.21$^{+0.65}_{-0.91}$   \\
 IG\_17\_P1\tablefootmark{b}   &   ---	 & --- &  ---              & ---                  & --- & --- \\
 IG\_51\_P3\tablefootmark{c}  &   ---    & 35.781 &  ---              & ---                  & ---& --- \\
 IG\_85\_P6   &  36.96  & 36.61 &  -0.05 $\pm$ 0.14 & 16.98$^{+40}_{-14}$   & 0.12$^{+0.39}_{-0.00}$ & 0.09$^{+0.36}_{-0.08}$  \\
 IG\_52\_P7   &  36.69  & 36.28 &  -0.21 $\pm$ 0.17 & 2.14$^{+16}_{-1}$     & 0.004$^{+0.06}_{-0.001}$ & 0.005$^{+0.05}_{-0.00}$   \\
 IG\_23\_P8   &  37.11  & 36.93 &  0.35  $\pm$ 0.19 & 207.54$^{+50}_{-111}$ & 5.51$^{+4.66}_{-4.18}$ & 4.92$^{+5.48}_{-3.73}$  \\
\hline                  
\end{tabular}
\tablefoot{
\tablefoottext{a}{Age from FUV-NUV and STARBURST99.}
\tablefoottext{b}{Regions with no emission neither in FUV nor NUV-band.}
\tablefoottext{c}{Regions with no emission in FUV band.}
}
\end{table*}

\section{Discussion}

Rings of neutral hydrogen around the galaxies are not very common in
the nearby universe. The mechanism which is able to produce such rings
is still not clear. One of the most famous systems is the Leo's H{\sc
  I} ring with a diameter $\sim$200 kpc, a M$_{HII}$
$\sim$10$^9$M$_\odot$ \citep{Schneider89b} and a density from
2$\times$10$^{18}$ to 6.4$\times$10$^{19}$ \citep{Schneider89a}, quite
asymmetric and somewhat clumpy. This system is similar to the H{\sc I}
ring around NGC~2865, but differs in the central galaxies: while the
Leo ring circles two galaxies, a spherical elliptical, M 105 and a S0
galaxy, NGC~3384, with no strong disturbances \citep{Michel10}, NGC~2865 is a shell galaxy. Another system with similar
  characteristics is described by \cite{Bettoni10} who found an
  extended outer ring of atomic gas around the lenticular galaxy
  NGC~4262. For both these systems, faint UV sources are detected in
  association with the H{\sc I} rings and a low star formation rate is
  estimated, $\sim$10$^{-3}$ M$_\odot$ yr$^{-1}$ \citep[][, Leo's and
    NGC~4262, respectively ]{Thilker09,Bettoni10}.

\subsection{Intergalactic H{\sc II} regions}

In this work we found seven intergalactic H{\sc II} regions around the
galaxy NGC~2865, at the same positions of H{\sc I} debries of low
density. The ages of the stellar populations associated with these
emissions are young ($<$ 200 Myr) and with masses no larger than
17$\times$10$^{6}$ M$_\odot$. Given these masses and ages, all regions
found in this work are considered as young star forming regions (or
clusters). Our results are in agreement with previous work by
\cite{Knierman03}, who found that structures formed in tidal tails
(optical and H{\sc I}) may manifest themselves either as clusters
along the tail or as larger systems such as dwarf galaxies, but not in
both. Note, however that the H{\sc II} regions found in our work
belong to tidal tails seen in H{\sc I}, not in
optical. \citet{Gerhard02} also detected an H{\sc II} region in a low
density H{\sc I} debris in the Virgo cluster, with an upper limit to
the density of 1$\times$10$^{19}$ cm$^{-2}$
\citep{Oosterloo05}. Recently, \citet{Yagi13} identified four
star-forming systems in the tail of NGC~4388 in the Virgo
cluster. Other authors reporting similar results are \citet{mendes04}
whom reported spectroscopic confirmation of four intergalactic H{\sc
II} region in the HCG92 and \citet{ryan04} whom found four isolated
H{\sc II} regions in two systems, NGC~1533 and HCG~16. All these
regions are located in low density of HI $\sim$ 10$^{19-20}$
cm$^{-2}$, similar to that found in this work.

H{\sc II} regions with similar masses and ages were found by
\citet{torres12} for the system NGC~2782 and \citet{demello12} for
HCG~92. Interestingly, these regions are withim H{\sc I} density
contours, but they are not superposed with any H{\sc I} peaks. This is
different from the case of tidal dwarf galaxies (TDGs), which are
expected to be found at the location of H{\sc I} peaks, like HCG~100
\citep{demello12}. Hence it seems that the formation of TDGs is not
favored in the environment around NGC~2865, because of the absence of
significant overdensities in the H{\sc I} distribution.

The very faint continuum emission of the seven intergalactic H{\sc II}
regions, similarly to other star clusters found in interacting
systems, suggests that these stellar populations are the first to be
formed and there were only very few stars or none previously. More
data on these star clusters, especially in the near infrared to
constraint intrinsic absorption, are necessary to confirm this.

\subsection{Metallicity}

The metallicity of the elliptical galaxy NGC~2865 is nearly
solar \citep{Hau99}, which is comparable with the metallicity of
galaxies whose morphological types are earlier than Sc. If NGC~2865 is
a merger remnant of two disk spirals, then there might be some
pre-enriched gas around this system, which was dispersed 
in the intracluster medium during such an event. 

If such pre-enriched material exists, then the young star forming
regions that form out of it would not follow the correlation
between luminosity and metallicity for classical dwarf galaxies
\citep{Skillman89}. They would have a luminosity that corresponds to
local star formation and a relative high metallicity inherited from
their progenitor galaxies.  The break in the correlation between
luminosity and metallicy is a diagnostic to separate `classical' dwarf
galaxies and young star clusters formed in galaxy interactions.
Hence the measurements of the metallicity of the young star forming regions
around NGC~2865 can provide the evidence for pre-enriched gas.

Indeed, the presence of such pre-enriched gas is supported by the
value for the metallicity measured for the intergalactic H{\sc II}
region IG\_04, of about $12+log(O/H)=8.7\pm0.18$, which is very near
to the solar value. Detections of young star forming regions with
solar metallicity is not uncommon: about ten star-forming regions are
reported in the literature. For example \cite{Duc00} measured
one such region in Arp~245; \cite{mendes04} measured four star forming
regions with metallicity near solar in the compact group HCG~92;
\cite{Michel10} found one region in the Leo Ring; \cite{demello12}
detected two regions in HCG~100; \cite{torres12} measured seven
regions in NGC~2782. Recently \cite{Yagi13} detected four regions in
the Virgo cluster also. The detection of young star forming regions
with metallicity (12 + log(O/H)) larger than 8.3 indicate that merger
events may be responsible for enriching the intergalactic medium with
metals by dispersing pre-enriched gas from the progenitor galaxies, also observed in the galaxy NGC~92 by \cite{Torres14}.
Also the young ages estimated for all the H{\sc II} regions around NGC~2865 may indicate that they formed in situ and from pre-enriched gas.

The comparison between the metallicity of IG\_04 and FGCE~0745 (12 + log(O/H) = 8.7 and 8.0, respectively)
indicates that the gas out of which the stars are forming in
IG\_04 has a different origin from that in the satellite
galaxy. Rather, its origin is most likely linked to NGC~2865 or the
progenitor galaxies which merged to form NGC~2865. Additional deeper
spectroscopic follow-up for the remaining six regions is required to
detect the [N{\sc II}] lines independently and confirm the presence of
pre-enriched gas in these star-forming regions also.

\subsection{The satellite galaxy FGCE 0745}

With the MSIS technique we measured the H$\alpha$ emissions in the
satellite galaxy FGCE 0745, in addition to those emissions coming from
the intergalactic H{\sc II} regions. We identify nineteen H$\alpha$
emitters in this galaxy. From the observed wavelength of the
H$\alpha$ lines, we measured the heliocentric line-of-sight velocities
in the range from 2747 to 3086 km s$^{-1}$, and a systemic velocity
for the galaxy center of 2920 km s$^{-1}$. From the H{\sc I} gas,
\citet{schiminovich95} reports that the systemic velocity of FGCE~0745
is $\sim$ 2725 km s$^{-1}$, with is somewhat bluer by $\simeq 200$ km
s$^{-1}$ than our estimated velocity.

We are also able to derive a velocity field and the rotation
curve, which are shown in Figure~\ref{velocity_map} and
\ref{rot_curve}. Assuming circular orbits for the ionized gas and the
centrifugal equilibrium, we estimated a total mass of $2.5 \times 10
^{10}$ M$_\odot$. From the Tully-Fisher relation \citep{Torres11} and
V$_{max}$ $\sim$ 160 km s$^{-1}$, we estimated an approximate stellar
mass of $1.2 \times 10 ^{10}$ M$_\odot$ and a total baryonic mass
(including the H{\sc I} gas) of 1.4 $\times$10$^{10}$.

From the total H$\alpha$ luminosity, we estimated a lower limit
to the total SFR of $2.3 \times 10^{-2}$ M$_\odot$ yr$^{-1}$: this
value is consistent with those measured in few low mass galaxies in
compact groups \citep{Bitsakis11}.

\section{Conclusions}

With the Multi-Slit Imaging Spectroscopy (MSIS) technique and the
Gemini telescope, we perform a flux limited H$\alpha$ survey in a
5\arcmin $\times$ 5\arcmin\ field centred on the southern part of the
H{\sc I} ring circling the elliptical galaxy NGC~2865. We detect
nineteen H$\alpha$ emissions within the satellite galaxy FGCE~0745 and
seven intergalactic H$\alpha$ emitters on the southern part of the
H{\sc I} ring. The line-of-sight velocities of the H$\alpha$
emissions are consistent with the systemic velocity and central
velocity dispersion of NGC~2865 and the satellite galaxy FGCE~0745.

Using the H$\alpha$ fluxes, we derive a constant star formation rate
of 2.6$\times$10$^{-3}$ M$_\odot$ yr$^{-1}$ for the entire sample of
intergalactic H{\sc II} regions. Such a low SFR is consistent with
star formation occurring in a very low H{\sc I} density environment.

We measured GALEX FUV and NUV fluxes for five out of the seven
intergalactic H{\sc II} regions around NGC~2865.  The color of these
sources are consistent with a wide range of ages ($<$200My) and masses
(1.7$\times$10$^{7}$ - 5.0$\times$10$^{3}$ M$_\odot$). The young ages
and the low masses suggest that these star clusters were formed in
situ out of gaseous debris at those locations.

We are also able to estimate the metal abundance of one intergalactic
H{\sc II} region, IG\_04, to be 12+log(O/H)= 8.7, which is similar to
the value determined for NGC~2865. Such metallicity value is higher
that the one expected for a dwarf galaxy of similar luminosity as the
intracluster region IG\_04.

We also detect the H$\alpha$ emissions associated with the satellite
galaxy FGCE~0745. We identify nineteen sources along the galaxy major
axis and determine a maximum rotation velocity of V$_{max} \simeq$ 160
km s$^{-1}$.  By coadding the spectra from the H{\sc II} emission in
FGCE~0745, we are able to measure the [N{\sc II}]$\lambda$6548 line,
and derive the metallicity for this satellite galaxy, which is lower
than solar.

The comparison between the metallicity for the gas in the satellite
galaxy and that for the intergalactic region IG\_04 indicate that the
gas in IG\_04 has a different origin, most likey it comes from gaseous debris
dispersed in the intracluster regions by the merging event that led to the
formation of  NGC~2865.

Deeper spectroscopy of the intergalactic sources and the NGC~2865 galaxy
are required to further explore and determine the internal extinction
of these star forming regions and place stronger constraints on their
ages and metal content.

\begin{acknowledgements} The authors would like to thank the anonymous
referee for the thoughtful comments which improved the clarity of this
paper. We would like to thank C. D. P. Lagos, L. Coccato and
A. Longobardi for help and useful discussions. This work is based on
observation obtained with the Gemini telescope. This work has made
used of NED-database, Mikulski Archive for Space Telescope
(MAST). FU-V acknowledges the financial support of ESO studentship
programme and CNPq through an PhD. CMdO thanks FAPESP for funding
through thematic grant 2006/56213-9. ST-F acknowledges the financial support of the Chilean agency FONDECYT through a project ``Iniciaci\'on en la Investigaci\'on'', under contract 11121505 and ST-F also acknowledges the support of the project CONICYT PAI/ACADEMIA 7912010004. DdFM was partially funded by NASA ADAP NNX09AC72G.
  \end{acknowledgements}

\bibliographystyle{aa} 
\bibliography{urrutia} 

\begin{appendix} 

\section{Spectra FGCE07445}\label{app:A}

We present the spectra for the H$\alpha$ sources inside the satellite galaxy FGCE 0745. In total we found 19 H$\alpha$ emitting regions and for all of them the H$\alpha$ is clearly distinguishable. The identification of the regions are on the top-left of each panel. The principal parameters derived from the H$\alpha$ line are in Tables  \ref{table:basic_dwarf} and \ref{table:Dwarf}.
   \begin{figure}
   \centering
   \includegraphics[width=\hsize]{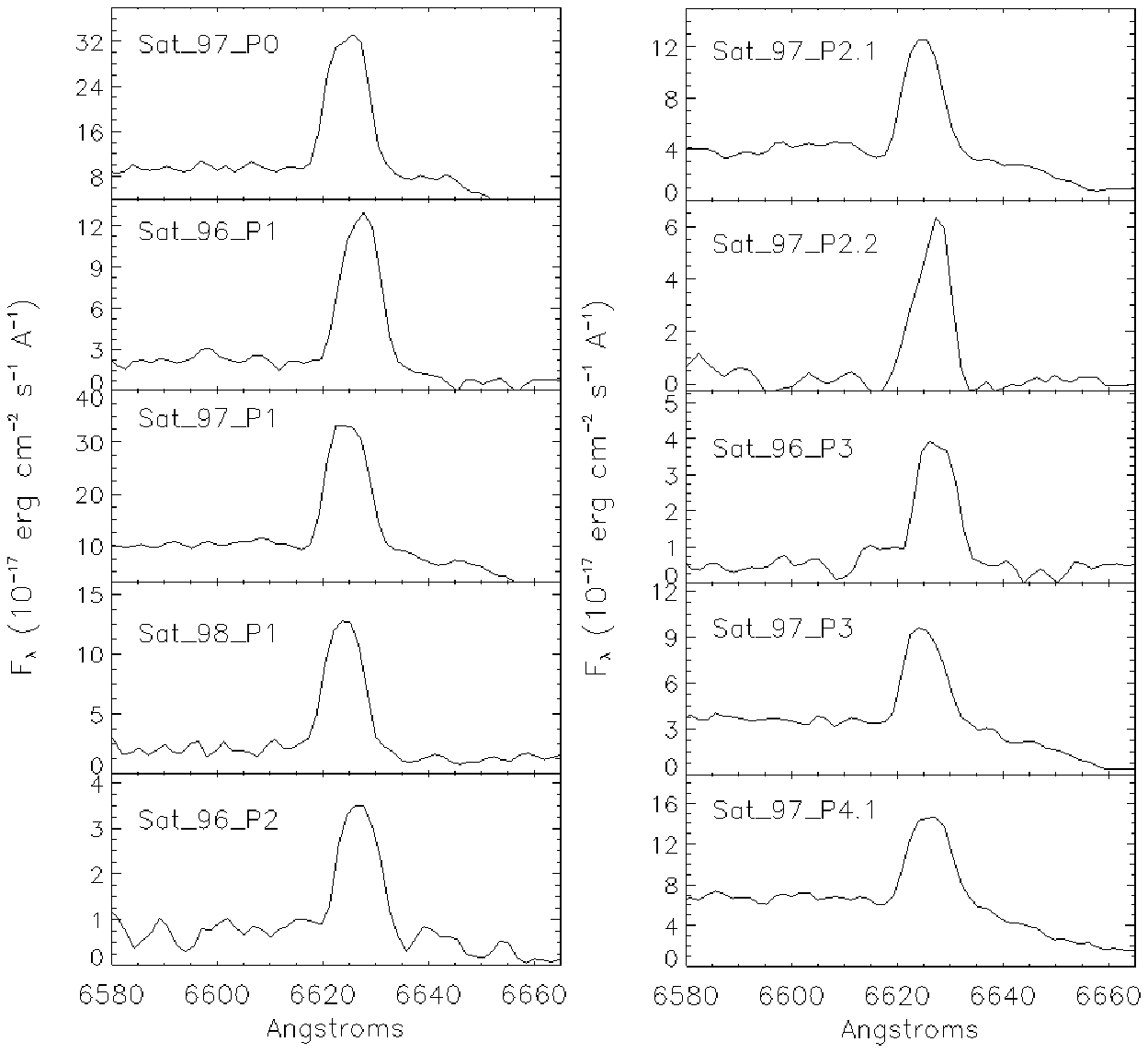}
         \includegraphics[width=\hsize]{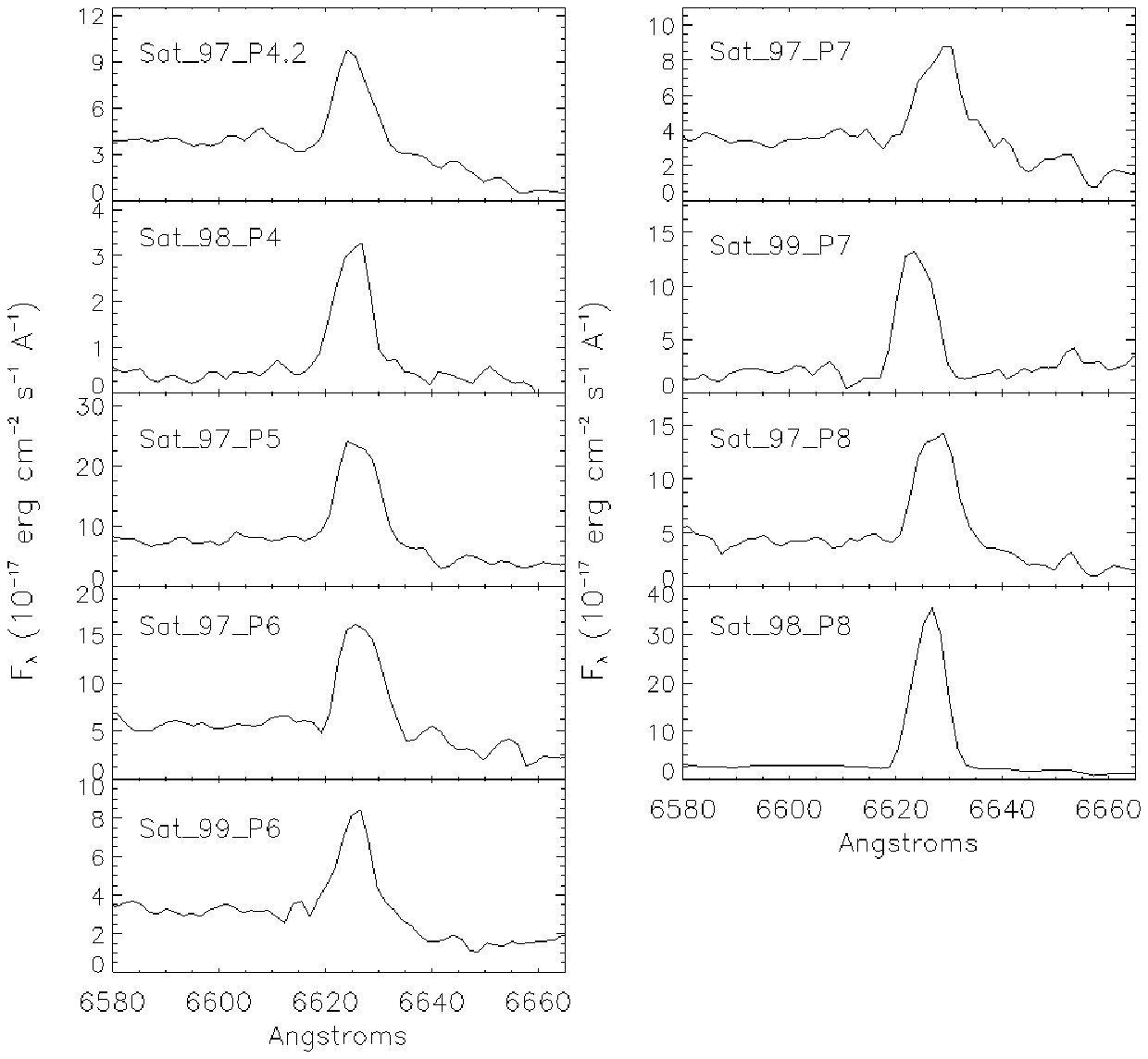}
      \caption{Spectra of the H$\alpha$ emitter regions detected in the satellite galaxy FGCE 0745.}
         \label{spectra_dwarf21}
   \end{figure}

\end{appendix}

\end{document}